\documentclass[preprint,12pt,showpacs,showkeys,preprintnumbers,pre,amsmath,amssymb]{elsarticle}

\usepackage{amssymb}
\usepackage{amsmath}
\usepackage[usenames,dvipsnames,x11names,table]{xcolor}
\usepackage{hyperref}
\usepackage{graphicx}
\usepackage{mathtools}
\usepackage[normalem]{ulem}
\usepackage{amsthm}
\usepackage[capitalize]{cleveref}

\usepackage{makecell}

\usepackage{tabularx}

\newcounter{bla}

\usepackage{color}

\journal{Computer Physics Communications}

\begin{document}

\begin{frontmatter}



\title{TLBfind: a Thermal Lattice Boltzmann code for concentrated emulsions with FINite-size Droplets}


\author[a]{Francesca Pelusi}
\author[b]{Matteo Lulli\corref{author}}
\author[c]{Mauro Sbragaglia}
\author[d]{Massimo Bernaschi}

\cortext[author] {Corresponding author.\\ \textit{E-mail address:} lulli@sustech.edu.cn}
\address[a]{Helmholtz Institute Erlangen-Nürnberg for Renewable Energy, Forschungszentrum Jülich, 91058 Erlangen, Germany}
\address[b]{Department of Mechanics and Aerospace Engineering, Southern University of Science and Technology, Shenzhen, 518055 Guangdong, China}
\address[c]{Department of Physics \& INFN, University of Rome “Tor Vergata”, 00133 Rome, Italy}
\address[d]{Istituto per le Applicazioni del Calcolo (IAC) - CNR, 00185 Rome, Italy}

\begin{abstract}
In this paper, we present TLBfind, a GPU code for simulating the hydrodynamics of droplets along with a dynamic temperature field. TLBfind hinges on a two-dimensional multi-component lattice Boltzmann (LB) model simulating a concentrated emulsion with finite-size droplets evolving in a thermal convective state, just above the transition from conduction to convection. The droplet concentration of the emulsion system is tunable 
and at the core of the code lies the possibility to measure a large number of physical observables characterising the flow and droplets. Furthermore, TLBfind includes a parallel implementation on GPU of the Delaunay triangulation useful for the detection of droplets' plastic rearrangements, and several types of boundary conditions, supporting simulations of channels with structured rough walls.
\end{abstract}


\end{frontmatter}



\noindent {\bf PROGRAM SUMMARY}

\begin{small}
\noindent
{\em Program Title:} TLBfind \\
{\em CPC Library link to program files:} (to be added by Technical Editor) \\
{\em Developer's repository link:} \url{https://github.com/FrancescaPelusi/TLBfind} \\
{\em Code Ocean capsule:} (to be added by Technical Editor)\\
{\em Licensing provisions:} MIT  \\
{\em Programming language:} CUDA-C      \\
{\em Nature of problem:} Hydrodynamics of concentrated emulsions with finite-size droplets in a thermal convective state.\\
{\em Solution method:} Single relaxation time Lattice Boltzmann (LB) method to solve Navier-Stokes equations for fluids, coupled with the temperature field dynamics
. The output describes the dynamics of finite-size droplets of concentrated emulsions in presence of a temperature field. The temperature field obeys the advection-diffusion equation.\\
{\em Additional comments including restrictions and unusual features:} Plastic rearrangements of droplets are detected via the parallel implementation of the Delaunay triangulation, and boundary conditions are tunable.\\
   \\
\end{small}
%
\section{Introduction}\label{sec:intro}
Understanding the hydrodynamic behaviour of concentrated emulsions -- and in general soft particles suspensions --  represents an intriguing subject of study in the context of fluid dynamics, with a wide range of applications, from everyday life situations to modern technologies~\cite{Hetsroni82,Gallegos99,Khanetal11,Shao15,McClements15,Yukuyama16,Wang2019}. Although many questions have been answered in the last decades~\cite{PrincenKiss89,Schramm92,Barnes94,Coussot05,BarratReview17}, many other aspects still deserve further scrutiny, such as the precise understanding of the heat transfer properties in these materials when evolving in a thermal convective state. This may be relevant in a variety of situations including the convective motion of magma~\cite{Stein92} composed of a melt with crystal suspensions where convection may be limited to the hotter (less crystalline) portions, the oil recovery industrial processes~\cite{Zhou19} aiming at using emulsions as packer fluids in order to significantly reduce the heat transfer rate during extraction, the production of slurry ice~\cite{Egolf05}, i.e., ice crystals distributed in water or an aqueous solution typically used to replace ice and salt for food cooling. The need for numerical simulations allowing to study the convective heat transfer in model emulsions motivated the present work.\\
Emulsions are structurally characterised by a collection of “soft domains" (i.e., droplets) of a dispersed phase in another continuous phase. The emulsion droplets concentration influences the rheological properties of the system: dilute emulsions behave as Newtonian fluids, with the viscosity being constant regardless of the applied shear rate; more concentrated emulsions display a non-Newtonian mechanical response and exhibit a viscosity that depends non-linearly on the applied shear rate~\cite{Pal2000}. Furthermore, there exists a critical concentration above which emulsions can be categorised as yield-stress fluids~\cite{Barnes94,Larson,Bonn17}, wherein the system shows a stress threshold (the so-called yield stress) below which no flow is observed and above which the emulsion flows displaying shear-thinning. Emulsions have been characterised in the literature via many experimental studies (see~\cite{Kilpatrick12,Kale17,Mcclements18} for reviews). However, experimental insights may be arduous to capture at scales comparable with the droplet size. This issue warrants the use of numerical simulations to investigate the response of emulsions and, more generally, soft particle suspensions. For the latter, different numerical methods have been exploited, such as the boundary integral method~\cite{Rahimian10}, the discrete element method~\cite{Yang15,Kroupa16}, molecular dynamics simulations~\cite{Seth11,Mansard13,Jung21}, and lattice Boltzmann (LB) models~\cite{Dollet15,Derzsi17,Derzsi18} just to cite some examples. The focus of this paper is on LB models, which have become very popular and attractive in the last two decades, due to their simplicity, efficiency and applicability in different contexts~\cite{Kruger17,Succi18}. Several open-access LB implementations are available, e.g., Ludwig~\cite{Ludwig01}, LB3D~\cite{LB3D_2017}, LBsoft~\cite{LBsoft20}, Palabos~\cite{Palabos21}, and LBfoam~\cite{LBfoam21}. However, an LB software that simulates the dynamics of finite-size droplets stabilised against coalescence and subjected to thermal convection is not available to the best of our knowledge in an open-access version. We aim at filling this gap by presenting TLBfind, a code based on a two-dimensional LB scheme devised to simulate emulsions with finite-size droplets where the temperature field dynamics is coupled with the emulsion momentum equation.
We consider an emulsion confined between two parallel walls, heated from below and cooled from above, i.e., in the paradigmatic set-up of the Rayleigh-B{\'e}nard convection~\cite{Moore73,Benard1900,Rayleigh1916,Lohse10}. Studies on the thermal response of this kind of emulsions have been presented in a recent work~\cite{PelusiSM21}. TLBfind was tested to be optimal to study the heat transfer properties in the regime where the system sustains a convective state, just above the transition from conduction to convection.
In TLBfind it is possible to tune the droplet concentration, allowing to systematically transition from diluted to concentrated emulsions. Further flexibility of TLBfind lies in a variety of different boundary conditions that can be explored, and the high degree of parallelisation obtained on GPU using the CUDA-C language entails a huge efficiency, by saving computational costs and time.\\
The paper is organised as follows: in Section~\ref{sec:LBM} we briefly explain the LB scheme implemented in TLBfind. In Section~\ref{sec:testFlat} we give a test case of the Rayleigh-B{\'e}nard convection in concentrated emulsion when the walls of the channel are flat; a test case with rough walls will be analysed in Section~\ref{sec:testRough}, then we summarise the potentialities of TLBfind in Section~\ref{sec:conclusions}.
\section{Method}\label{sec:LBM}
We leverage an LB method~\cite{Kruger17,Succi18} for non-ideal multi-component systems that allows the mesoscopic simulation of a collection of finite-size droplets evolving in a thermal convective state. The isothermal counterpart of the model has already been presented and validated in many publications~\cite{Benzietal09,Sbragagliaetal12,Benzietal14,Dollet15,LinLin18,PelusiSM19,PelusiEPL19}. TLBfind considers the advection-diffusion dynamics of a scalar temperature together with the multi-component model. We summarise the main essential features of the methodology, by treating separately the isothermal multi-component fluid and the temperature dynamics.
%
%
\subsection{Multi-component lattice Boltzmann}\label{subsec:multiLBM}
\begin{figure}[t!]
\centering
\begin{tabular}{c}
\includegraphics[width=.62\linewidth]{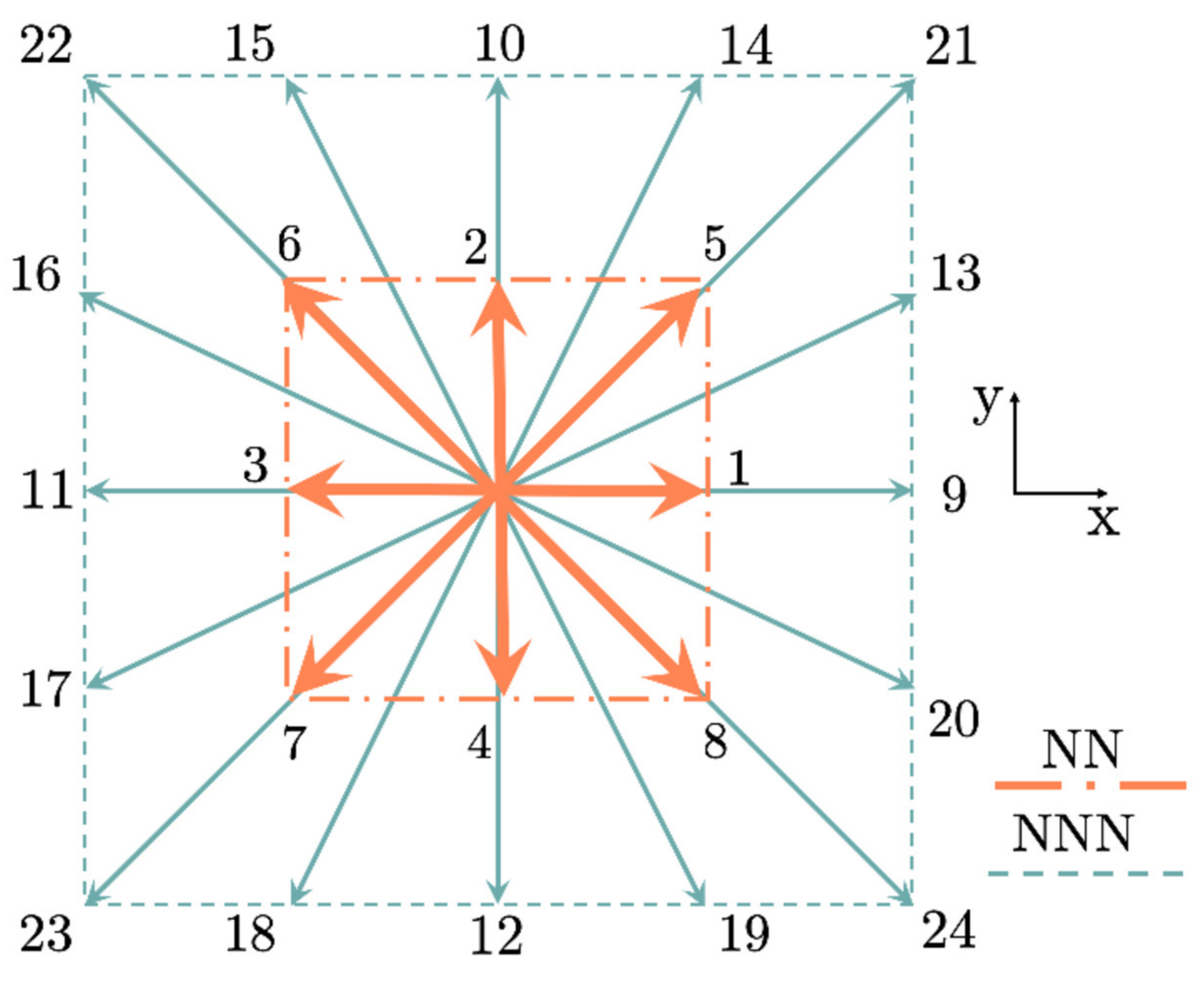}\\
\small (a) \\
\\
\begin{tabular}{|| c c c c c ||}
\hline
$w_i=w(\bf{c}_i)$ & $p_i=p(\bf{c}_i)$  & $|\bf{c}_i|$ & $\bf{c}_i$ & $i$ \\
\hline
4/9 & 247/420  & 0 & (0,0) & 0\\
1/9 & 4/63 & 1 & $(\pm1,0)$;$(0,\pm 1)$ & 1-4\\
1/36 & 4/135 & 2 & $(\pm 1, \pm 1)$ &5-8\\
0 & 1/180 & 4 & $(\pm 2,0)$ ; $(0,\pm 2)$ & 9-12\\
0 & 2/945 & 5 & $(\pm 1, \pm 2)$; $(\pm 2; \pm 1)$ & 13-20\\
0 & 1/15120 & 8 & $(\pm 2,\pm 2)$ & 21-24\\
\hline
\end{tabular}\\
\\
\small (b) \\
\end{tabular}
\caption{Panel (a): a scheme of the two neighbours sets of discrete velocities used for the evaluation of the interaction forces: a “nearest-neighbours" (NN) and a “next-to-nearest neighbours" (NNN) zone.
The NN zone corresponds to the set of velocities ${\bf c}_i$ of the D2Q9 LB method used, with index $i=0...8$. NNN are links with index $i = 9...24$. Concerning the interactions, each fluid component self-interacts via competing interactions, i.e., attractive forces involving the NN (Eq.~\eqref{eq:FintAttrattive}), and repulsive ones (Eq.~\eqref{eq:FintRepulsive}) acting on both zones. On the other hand, the fluid-fluid cohesion interactions Eq.~\eqref{eq:FAB} are given by summing on sites in the NN zone.
Panel (b) shows the list of weights appearing in Eqs.~\eqref{eq:FAB},~\eqref{eq:FintAttrattive} and~\eqref{eq:FintRepulsive}.\label{fig:LBM}}
\end{figure}

We consider a two-dimensional system consisting of two components (labelled by $\ell = 1,2$). The LB dynamics considers mesoscopic probability density functions $f_{\ell,i}({\bf x},t)$, representing the fluid particle (mass) density of the component $\ell$ at the space-time location $({\bf x},t)$ and velocity ${\bf c}_i$, where the position $\mathbf{x}$ takes on the integer values of the nodes coordinates in a squared lattice. The $f_{\ell,i}({\bf x},t)$'s evolve via the discrete lattice Boltzmann equations:
\begin{equation}\label{eq:LBM}
f_{\ell,i}({\bf x}+{\bf c}_i,t+1) - f_{\ell,i}({\bf x},t) = \Omega_{\ell,i} ({\bf x},t),
\end{equation}
where both the lattice spacing $\Delta x$ and time step $\Delta t$ are considered equal to the unity. The r.h.s. of Eq.~\eqref{eq:LBM} encodes the physical effects of collisions, i.e., to redistribute particles among the populations $f_{\ell,i}({\bf x},t)$ at each node. Collisions are implemented via the Bhatnagar–Gross–Krook (BGK) operator~\cite{BGK54}, approximating the relaxation of $f_{\ell,i}({\bf x},t)$ towards the local equilibrium distribution $f_{\ell,i}^{(eq)}$ with a relaxation time $\tau$:
\begin{equation}\label{eq:BGK}
\Omega_{\ell, i} ({\bf x},t) = -\frac{1}{\tau} \left[f_{\ell,i}({\bf x},t)-f_{\ell,i}^{(eq)}\left(\rho_{\ell}({\bf x},t),{\bf u}_\ell ({\bf x},t)\right) \right].
\end{equation}
The local equilibrium $f_{\ell,i}^{(eq)}$ is the Maxwellian distribution, given in the form
\begin{equation}
f_{\ell,i}^{(eq)}\left(\rho_{\ell},{\bf u}_\ell \right)=w_i \rho_{\ell} \left[1+\frac{u_{\ell,k} c_{ik}}{c_s^2}+\frac{u_{\ell,k} u_{\ell, p} (c_{ik} c_{ip}-c_s^2 \delta_{kp})}{2 c_s^4} \right],
\end{equation}
where the $w_i$ are the lattice weights, $c_s^2=\sum_i w_i |{\bf c}_i|^2 / d = 1/3$ is the squared sound velocity (with $d = 2$ the space dimension), and repeated indices are summed upon. The index $i$ runs on a finite number of values depending on the LB model used. In TLBfind, we leverage the widely used D2Q9 scheme (see Fig.~\ref{fig:LBM}(a), orange arrows), with $9$ lattice velocities (i.e., $i=0,\ldots,8$) in a two-dimensional domain. The associated weights $w_i$ are reported in Fig.~\ref{fig:LBM}(b). After the collision, the l.h.s. of Eq.~\eqref{eq:LBM} performs a streaming of each fluid particle (mass) density on the lattice following the set of lattice velocities ${\bf c}_i$. 
Information on coarse-grained fields of densities of each component ($\rho_{\ell}$) and emulsion velocity (${\bf u}$) can be extracted by by taking moments in the discrete velocity space as
\begin{equation}
\rho_{\ell}({\bf x},t)  =\sum_i f_{\ell,i} ({\bf x},t),
\end{equation}
\begin{equation}\label{eq:velo}
{\bf u} ({\bf x},t)  =\frac{1}{\rho}\sum_{\ell,i} {\bf c}_i f_{\ell,i}({\bf x},t),
\end{equation}
where we used the total density defined as $\rho=\sum_{\ell} \rho_{\ell}$.

\noindent The effects of interaction forces, ${\bf F}^{\mbox{\tiny int}}_{\ell}({\bf x},t)$, and external volume forces, ${\bf F}^{\mbox{\tiny ext}}_{\ell}({\bf x},t)$ (the latter will be discussed in Section~\ref{subsec:thermLBM}), are encompassed in a source term ${\bf F}_{\ell}({\bf x},t) = {\bf F}^{\mbox{\tiny int}}_{\ell}({\bf x},t) + {\bf F}^{\mbox{\tiny ext}}_{\ell}({\bf x},t)$ that enters in Eq.~\eqref{eq:BGK} as a shift in the hydrodynamic velocity~\eqref{eq:velo}:
\begin{equation}\label{eq:shift}
{\bf u}_\ell ({\bf x},t) = {\bf u}({\bf x},t) + \dfrac{\tau {\bf F}_{\ell}({\bf x},t)}{\rho_\ell}.
\end{equation}
The interaction forces ${\bf F}^{\mbox{\tiny int}}_{\ell}({\bf x},t)$ hinge on the multi-range Shan-Chen methodologies~\cite{ShanChen93,Shan94,Falcucci07,Falcucci10} and include three different contributions for the case presented in this paper~\cite{Benzietal09, Sbragaglia07}: ${\bf F}^{x}_{\ell}({\bf x},t)$,  ${\bf F}^{a}_{\ell}({\bf x},t)$, and ${\bf F}^{r}_{\ell}({\bf x},t)$. The term ${\bf F}^{x}_{\ell}({\bf x},t)$ represents the interactions between fluid elements of the two different species; it is introduced with the aim of creating phase segregation and the formation of stable diffuse interfaces, i.e., interfaces characterised by a finite width. In formulae, it reads:
\begin{equation}\label{eq:FAB}
{\bf F}^{x}_{\ell} ({\bf x},t) = - {\cal G}_{12} \psi_{\ell}({\bf x},t) \sum_{\ell', \ell' \ne \ell}\sum_{i \in \text{NN}} w_i \psi_{\ell'}({\bf x}+{\bf c}_i,t) {\bf c}_i,
\end{equation}
where $\psi_{\ell}({\bf x},t)=\psi(\rho_{\ell}({\bf x},t))$ is the pseudo-potential function of the Shan-Chen formulation~\cite{ShanChen93}, and ${\cal G}_{12}$ is a positive coupling constant dictating the strength of these interactions~\footnote{${\cal G}_{12}$ also controls the width of the interface.}. In the specific case, we use the simplest pseudo-potential, i.e.,
\begin{equation}
\psi_{\ell}({\bf x},t)=\dfrac{\rho_\ell}{\rho_0}
\end{equation}
where $\rho_0$ is a reference density. The set NN refers to the set of “nearest neighbours" nodes of ${\bf x}$ on the lattice, which coincide with the set of the D2Q9 directions used for the streaming of the LB populations (see Fig.~\ref{fig:LBM}(a)). The effect of ${\bf F}^{x}_{\ell}({\bf x},t)$ is a change in the bulk pressure, which now features ideal contributions summed to the non-ideal ones:
\begin{equation}\label{eq:Pbulk}
P_b(\rho_1,\rho_2) = \underbrace{c_s^2 \rho_1 + c_s^2 \rho_2}_\text{ideal} + \underbrace{c_s^2 {\cal G}_{12}\rho_1 \rho_2}_\text{non-ideal}.
\end{equation}
The two other contributions to ${\bf F}_\ell^{\mbox{\tiny int}}$, i.e., ${\bf F}^{a}_{\ell}({\bf x},t)$ and ${\bf F}^{r}_{\ell}({\bf x},t)$, represent short-range attractive ($a$) and long-range repulsive ($r$) competing interactions, respectively~\cite{Benzietal09,LinLin18}. They are introduced to simulate the action of a positive disjoining pressure at the interface that inhibits the coalescence of droplets~\cite{Sbragagliaetal12} (see Fig.~\ref{fig:preparation}). In formulae, they read~\cite{Benzietal15}:
\begin{align}\label{eq:FintAttrattive}
{\bf F}^a_{\ell} ({\bf x},t)= & - {\cal G}^a_{\ell \ell} \psi_{\ell} ({\bf x},t) \sum_{i \in \text{NN}} w_i \psi_{\ell}({\bf x}+{\bf c}_i,t) {\bf c}_i \\
{\bf F}^r_{\ell} ({\bf x},t)= & - {\cal G}^r_{\ell \ell}\psi_{\ell} ({\bf x},t) \left[\sum_{i \in \text{NN}} p_i \psi_{\ell} ({\bf x}+{\bf c}_i,t) {\bf c}_i + \sum_{i \in \text{NNN}} p_i \psi_{\ell} ({\bf x}+{\bf c}_i,t) {\bf c}_i  \right], \label{eq:FintRepulsive}
\end{align}
with ${\cal G}_{\ell \ell}^a < 0$ and ${\cal G}_{\ell \ell}^r > 0$. In Eqs.~\eqref{eq:FintAttrattive}-\eqref{eq:FintRepulsive}, the pseudo-potential is used in the original Shan-Chen form~\cite{ShanChen93}:
\begin{equation}
\psi_{\ell}({\bf x},t) = \rho_0 \left[ 1- \exp (-\rho_\ell ({\bf x},t) /\rho_0) \right].
\end{equation}
The set of nodes NNN in~\eqref{eq:FintRepulsive} refers to “next-to-nearest neighbours", i.e., an additional layer of 16 lattice velocities beyond the D2Q9 links (see Fig.~\ref{fig:LBM}(a), light-blue arrows). The values of the weights $p_i$ are reported in Fig.~\ref{fig:LBM}(b).\\
By summing over the components $\ell$, the total force ${\bf F}({\bf x},t) = \sum_\ell {\bf F}_\ell$ acting on the momentum density (Eq.~\eqref{eq:shift}) is obtained.\\
\noindent The reference hydrodynamic equations at large scales are the diffuse-interface Navier-Stokes equations:
\begin{equation}\label{eq:NS}
\rho \left(\partial_t+u^{\mbox{\tiny (H)}}_k \partial_k\right) u^{\mbox{\tiny (H)}}_i  = -\partial_j P_{ij} + \eta_0 \partial_j \left(\partial_{i} u^{\mbox{\tiny (H)}}_j+\partial_{j} u^{\mbox{\tiny (H)}}_i\right) +f^{\mbox{\tiny ext}}_{i},
\end{equation}
where $\rho {\bf u}^{\mbox{\tiny (H)}}=\rho {\bf u}+{\bf F}/2$ is the hydrodynamical momentum density, and $i=x,y$. Notice that the hydrodynamic velocity ${\bf u}^{\mbox{\tiny (H)}}$ differs from Eq.~\eqref{eq:velo} because of forcing renormalisations. The bare viscosity $\eta_0$ in the hydrodynamic equations is related to the relaxation time $\tau$ of the LB equation~\eqref{eq:LBM} as:
\begin{equation}\label{eq:VISCO}
\eta_0=\rho c_s^2 \left(\tau-\frac{1}{2}\right),
\end{equation}
hence it can be tuned by changing the $\tau$ in the LB dynamics. The viscosity $\eta_0$ represents the “bare" viscosity, thus the bare viscous stress tensor $\eta_0  \left(\partial_{i} u^{\mbox{\tiny (H)}}_j+\partial_{j} u^{\mbox{\tiny (H)}}_i\right)$ is supplemented by the pressure tensor $P_{ij}$ that depends on the density heterogeneities~\cite{Dollet15}. These two contributions of stress sum up to give the total stress that is used in the rheological characterisations of the emulsions (see Section~\ref{subsec:rheology}). Finally, the term $f^{\mbox{\tiny ext}}_{i}$ in Eq.~\eqref{eq:NS} appears on behalf of the density of external forces (the buoyancy force density will be discussed in Section~\ref{subsec:thermLBM}).

\subsection{Boundary conditions and rough walls}\label{subsec:BC}
TLBfind allows for the exploration of both fully periodic and confined systems. Rough-wall-flags -- one associated with each wall -- are introduced with the aim of switching on/off the presence of roughness at the walls (these features will be discussed in detail in Sections~\ref{sec:testFlat} and~\ref{sec:testRough}). If the rough-wall-flags are switched off, we are in the presence of flat walls and the code implements a modified mass-conserving bounce-back rule~\cite{Succi18,Kruger17}, 
that is a bounce-back dynamics designed to assign the desired input value to the hydrodynamical velocity at the walls
\begin{equation}
{\bf u}^{\mbox{\tiny (H)}}_x ({\bf x}_b,t)=u_{\mbox{\tiny w}}{\bf e}_x
\end{equation}
where ${\bf x}_b$ are the boundary nodes. This set-up in TLBfind is used when performing rheological experiments (see Section~\ref{subsec:rheology}) in a Couette channel, where each wall moves with its own velocity $u_{\mbox{\tiny w}}{\bf e}_x$ along the stream-flow direction $x$.
In the presence of structured rough walls, rough-wall-flags are switched on and TLBfind operates with simple half-way bounce-back boundary conditions, with $u_{\mbox{\tiny w}}{\bf e}_x = 0$.
\subsubsection{Wetting conditions}
In the presence of walls, both flat and rough, TLBfind allows handling the fluid-wall interactions, i.e., the wetting conditions. The implementation is very simple, regarding just the definition of density values of the two components at the wall: the code sets these densities (and the pseudo-potential accordingly) in the ghost nodes along the $y$ direction ($y = 0, ny+1$)~\footnote{The ghost nodes are introduced along both $x$ and $y$ directions as part of the lattice to facilitate the collision and streaming step close to the boundaries with the aim to supply the boundary nodes with otherwise missing populations.} and in the wall nodes as equal to the values $\rho_{\mbox{\tiny w, max}}$ and $\rho_{\mbox{\tiny w, min}}$, respectively. More details on how to perform simulations with different wetting conditions are provided in Section~\ref{sec:testFlat}.
\subsection{Thermal lattice Boltzmann}\label{subsec:thermLBM}
An auxiliary probability distribution function $g_{i}({\bf x},t)$ is introduced with the purpose of simulating the dynamics of a scalar temperature field $T({\bf x},t)$. Similarly to the fluid populations $f_{\ell, i}$, $g_{i}({\bf x},t)$ are governed by a discrete lattice Boltzmann equation~\cite{Ripesi14,Succi18,Kruger17}
\begin{equation}\label{eq:LBMg}
g_{i}({\bf x}+{\bf c}_i,t+1) -g_{i}({\bf x},t)  = -\frac{1}{\tau_g} \left[g_{i}({\bf x},t)-g_i^{(eq)}\left( T({\bf x},t), {\bf u}^{\mbox{\tiny (H)}}({\bf x},t)\right) \right],
\end{equation}
and the temperature field is obtained by taking the moment of order zero of the distribution functions:
\begin{equation}
T({\bf x},t)=\sum_i g_{i}({\bf x},t).    
\end{equation}
The local equilibrium $g_i^{(eq)}$ for the temperature field takes the form~\footnote{The temperature field $T({\bf x},t)$ has to be interpreted as the relative temperature with respect to some reference value.}
\begin{equation}\label{eq:Teq}
g_i^{(eq)}(T, {\bf u}^{\mbox{\tiny (H)}})=w_i T \left[1+\frac{u^{\mbox{\tiny (H)}}_k c_{ik}}{c_s^2}+\frac{u^{\mbox{\tiny (H)}}_k u^{\mbox{\tiny (H)}}_p (c_{ik}c_{ip}-c_s^2 \delta_{kp})}{2 c_s^4} \right].
\end{equation}
The long-wavelength limit of~\eqref{eq:LBMg} approximates the advection-diffusion equation for the temperature field
\begin{equation}\label{eq:T}
\partial_t T + u^{\mbox{\tiny (H)}}_k \partial_k T = \kappa \partial_{kk} T,
\end{equation}
where the thermal diffusivity $\kappa$ is related to the thermal relaxation time $\tau_g$ as follows:
\begin{equation}\label{eq:K}
\kappa=c_s^2 \left(\tau_g-\frac{1}{2}\right).
\end{equation}
The buoyancy term in Eq.~\eqref{eq:NS} is defined as:
\begin{equation}\label{eq:BUOYANCY}
f^{(\mbox{\tiny ext})}_i = \rho \alpha g T \delta_{iy},
\end{equation}
with $\alpha$ being the thermal expansion coefficient, $g$ the gravity acceleration, and ${\bf e}_y$ the unit vector in the wall-to-wall direction. With the aim to obtain the buoyancy force in \eqref{eq:BUOYANCY}, we include in Eq.~\eqref{eq:shift} an external volume force that reads
\begin{equation}\label{eq:alphaG}
{\bf F}_{\ell}^{\mbox{\tiny ext}}({\bf x},t) = \rho_\ell ({\bf x},t) \, \alpha \, g \, T({\bf x},t) \, {\bf e}_y.
\end{equation}

\subsection{Delaunay triangulation}\label{subsec:delaunay}
%
\begin{figure}[t!]
\begin{center}
\begin{tabular}{c}
\includegraphics[width=1.\linewidth]{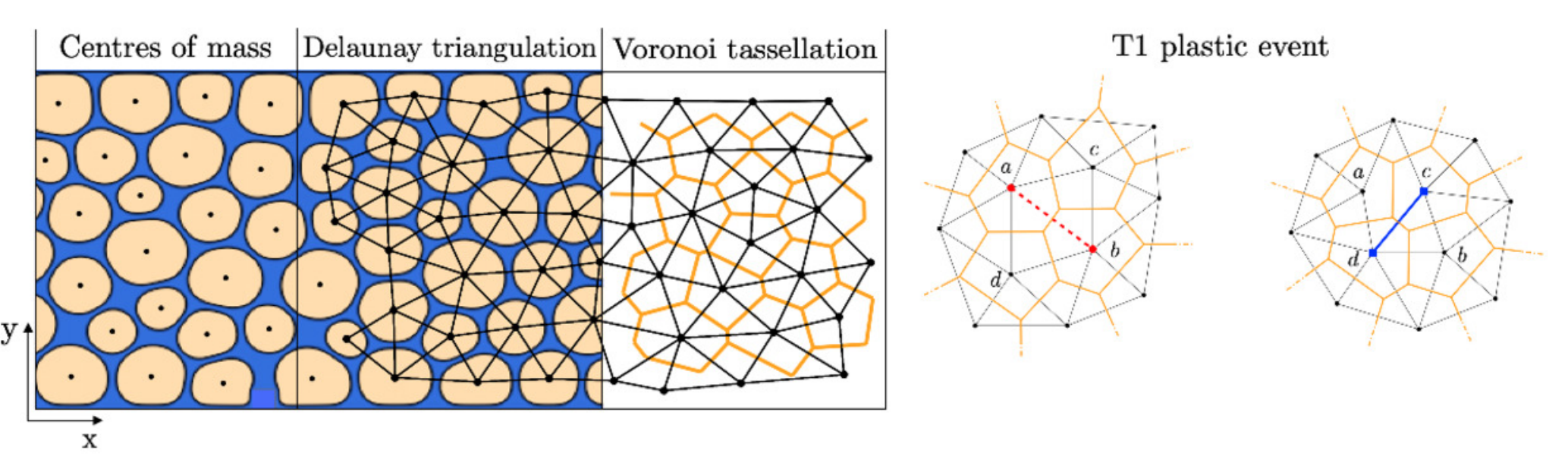} \\
\small (a)  \hspace{8cm}  \small (b) \\
\end{tabular}
\caption{Domains topology analysis of a two-dimensional concentrated emulsion. Dark-yellow regions refer to emulsion droplets and blue regions are occupied by the continuous phase/component. Panel (a): construction of Delaunay triangulation (black lines) and Voronoi tassellation (orange lines) based on the centres-of-mass positions (black dots). Panel (b): a sketch of detection of a T1 plastic rearrangement. The old (dashed red) link between droplets $a$ and $b$ goes to zero and the new (blue continuous) link is created.}\label{fig:delaunayVoronoi}
\end{center}
\end{figure}
TLBfind performs a run-time analysis of Delaunay triangulation~\cite{Delaunay_1934} on the centres of mass of the droplets. The main reason for executing the Delaunay triangulation on GPU is the performance improvement we obtain. In a very preliminary version of TLBfind we resorted to an existing CPU library to carry out the triangulation. However, that choice had two disadvantages: $i)$ we had to copy back data from the GPU to CPU and it is well know that the bandwidth between CPU and GPU is limited so that the copy is slow; $ii)$ the CPU library we used and other open-source alternatives we found were implemented without considering high-performance requirements. So we decided to implement our own, high-performance parallel implementation.  The Delaunay triangulation can be seen as an adjacency matrix of the droplets, storing information about the topology of the droplets arrangement. In particular, the triangulation is obtained in such a way that the circumference circumscribed at any of the triangles does not contain any other point of the set. A reference depiction is shown in Fig.~\ref{fig:delaunayVoronoi}. The algorithm implemented in TLBfind was described in detail in~\cite{Bernaschi16}, and here we report the main ingredients. The first analysis step consists in identifying the bulk of the droplets as those regions of the lattice where the density field of the first component is larger than a certain threshold $B_{th}$. Then, these sets of lattice points are identified by means of a GPU implementation of a clustering algorithm adapted from those used in accelerated Ising model simulations at the critical point~\cite{Swendsen_1987,Komura_2012}. We remark that, while in the case of the Ising model the construction of the cluster is probabilistic in nature, in our present case we use the algorithm in a \emph{geometric} way in order to identify all the nodes related to the bulk of a droplet.
This allows us assigning unique labels to all droplets and compute their centre of mass (black dots in Fig.~\ref{fig:delaunayVoronoi}$(a)$). The labels associated with each droplet are stored and employed to define a \emph{digital Voronoi} tessellation~\cite{Voronoi_1908} where each point of the lattice is “coloured" by the label of the nearest centre of mass (orange lines). In the continuum construction, a Voronoi tessellation is dual to a Delaunay triangulation, thus encoding the same information but in a metric language. The vertices at which adjacent digital Voronoi cells converge are used to identify the Delaunay triangulation (black links in Fig.~\ref{fig:delaunayVoronoi}$(a)$). Subsequent Delaunay triangulations are compared in order to detect T1 plastic rearrangements of the droplets, which have the effect of stably changing the topology of the droplets arrangement~\cite{Bernaschi16}. These topological events only involve 4 droplets, which can be thought of as being at the four vertices of a quadrilateral labelled $a$, $b$, $c$ and $d$, and they are associated with a "flip" of the diagonal link of the triangulation. For example, in Fig.~\ref{fig:delaunayVoronoi}$(b)$, the diagonal link between $a$ and $b$ (dashed red line), is replaced by the link between $c$ and $d$ (solid blue line). From the Voronoi tesselation perspective, at the degenerate configuration, i.e., when the four droplets all lie on a common circle, the Voronoi edge orthogonal to the Delaunay side shrinks to zero. This is an equivalent way, though less robust, to implement the T1 event detection. Avoiding dealing with such a metric structure makes the detection of plastic events via Delaunay triangulation more robust and easier to implement.

\subsection{GPU Implementation}\label{subsec:GPU}

One of the crucial requirements to achieve a good performance on the NVIDIA GPU
is that {\em global} memory accesses (both read and write) should be
coalesced. This means that memory access needs to be {\em aligned}
and coordinated within a group of threads. The basic rule is
that the thread with id $n\in\{0,\ldots,N-1\}$ should access element $n$ at byte address\\
$StartingAddress+sizeof(type)\cdot n$ where $sizeof(type)$ is
equal to either $4$, $8$ or $16$ and $StartingAddress$ is a
multiple of $16\cdot sizeof(type)$.

With this aim, the fluid populations of a lattice site are
not contiguous in the GPU global memory so that 
they are ordered following the {\em structure-of-arrays} data layout. 
All data not modified during the simulation, such as coefficients, are pre-computed during the
initialisation phase and stored in the GPU constant memory, which has
performances analogous to those of registers if, as in our case, all
the threads running on the same multiprocessor access the same
constant memory locations.

After the initialisation phase, all the computations required for the
LB update are performed on the GPU. A single step of the simulation is implemented through a sequence of CUDA kernels guaranteeing the correct sequential order
of the sub-steps. Each CUDA kernel implements a sub-step of the update
procedure ({\em e.g.}, collision, streaming)
by splitting the work among a configurable number of threads and blocks,
which may be fine-tuned to achieve optimal performance on different CUDA 
devices. Each thread works sequentially on a group of lattice nodes assigned to it.
For each lattice node, the thread copies data from the global memory
into registers, performs the computation and writes the results back
in the global memory. In order to manage the parallelisation of the
streaming phase without causing conflicts among multiple threads,
fluid populations are stored in the global memory using a ``double
buffer'' policy. At the end of the simulation, the
final results are copied back to the CPU main memory in order to be
saved on a secondary storage device. Through the input
file it is possible to require also the saving of partial results
of the simulation at regular intervals ({\em e.g.,} for check-pointing purposes).

Most of the global memory read-and-write operations are coalesced, with
the exception of a few reads relative to the computation
of the interaction forces and a few writes
relative to the streaming phase. In the first case, the calculation
of the force for a lattice node depends on values related to other
lattice nodes, which must be loaded from global memory even if alignment
requirements for coalesced accesses are not satisfied. In
the second case, target locations of the streaming phase are defined
by the lattice topology, and in general, they don't comply with the memory
alignment requirements.

For the function that computes the value of the hydrodynamic variables
all memory operations are ``local'', meaning that only the fluid
populations of a lattice site are required and that the resulting
hydrodynamic variables belong to the same lattice site. As a
consequence, there is not a single uncoalesced memory access.

Finally, as already mentioned, the fluid populations once uploaded on
the GPU memory do not need to be copied back to the main memory unless
a dump of the whole configuration is required. However, hydrodynamic
variables or other observables derived from them might be written back to the main memory much more
frequently since they represent the main physical output of the simulation.
Although the number of hydrodynamic variables per lattice site is
small compared to the number of fluid populations (there are 4
hydrodynamic variables {\em vs.} 9 fluid populations), so that the
run-time overhead of the copy from the GPU-memory to the CPU-memory is
small compared to the initialisation overhead, better performance may be obtained by reducing the number of these copy-back operations. Although the main goal of the present paper is to make TLBfind available and explain how to use it, we report some basic data about its computational performance. It is common to measure the performance of an LB code by using the LUPS (Lattice Updates per Second) metrics. TLBfind has, at least, two sets of populations for each fluid node and this means that both the amount of data to be moved from/to GPU global memory and the number of operations should be (at last) doubled. The number of MLUPS (Million of LUPS) that TLBfind is able to perform is $\sim 525$ on a Titan-V a GPU featuring 5120 CUDA cores (Volta architecture). This figure is perfectly in line with the performance (515 MLUPS) of {\em sailfish}~\cite{Januszewski_2014} an LB open-source code  (\url{https://github.com/sailfish-team/sailfish.git}) on the same hardware. By using Nsight, a user-friendly Nvidia visual profiler, we collected several additional information about the performance of TLBfind. In Table \ref{tab:metrics} we report two of them: the {\em warp} execution efficiency that shows a very high degree of utilisation and the throughput of the load global memory operations for three representatives kernels. For the Titan V the peak memory bandwidth is $\sim 650$ GB/s, and TLBfind achieves $> 80\%$ of that value. It is well known that LB codes are memory bandwidth bound meaning that the number of floating-point operations they execute is limited with respect to the amount of data that need to be loaded and stored from/to memory. This is true for both CPU and GPU codes and TLBfind is not an exception. As mentioned above, although TLBfind has, potentially, up to three sets of populations per node, the ratio between the number of arithmetic operations required by the execution of each LB time step and the amount of data moved from/to memory is fundamentally the same of any other ``simple'' 2D LB code. As to the Delaunay triangulation, we did not carry out specific performance measurements. However, we noticed that the Delaunay procedure takes a time that is, roughly, equivalent to two LB time steps.
Further details about the GPU implementation can be found in \cite{Bernaschi16} and \cite{BernaschiGPU09}.
\begin{table}[h!]
\begin{center}
\begin{tabular}{|p{6cm} p{5cm} p{2.5cm}|}
\hline
\rowcolor{GreenYellow} \multicolumn{3}{|c|}{CUDA performance metrics} \\
\hline
{\em kernel} & {\em metrics} & {\em average value} \\
\hline
\texttt{AVERAGE\_1} & Warp Execution Efficiency & \texttt{99.74\%}\\
\texttt{AVERAGE\_1} & Global Load Throughput & \texttt{548 GB/s}\\
\texttt{moveplusforcingconstructWW} & Warp Execution Efficiency & \texttt{95.6\% GB/s}\\
\texttt{moveplusforcingconstructWW} & Global Load Throughput & \texttt{524 GB/s}\\
\texttt{forcingconstructWW} & Warp Execution Efficiency & \texttt{99.85\%GB/s}\\
\texttt{forcingconstructWW} & Global Load Throughput & \texttt{538 GB/s}\\
\hline
\end{tabular}
\caption{Some CUDA performance metrics obtained by using Nsight.}\label{tab:metrics}
\end{center}
\end{table}
\section{Test case with flat walls}\label{sec:testFlat}

\noindent As a first example, we show a simulation of Rayleigh-B{\'e}nard convection of a model emulsion, by explaining in detail which are the parameters involved. With this aim, hereafter in the text, all words in Computer Modern Typewriter font are referred to as input parameters, and what will be defined as Boolean is considered turned off (false = 0) or on (true = 1). The same font will be used also for output files names. All tunable simulation parameters are listed in the input file \texttt{tlbfind.inp}.\\
We remark that, before starting a simulation, TLBfind requires a preparation step during which the structural properties of the system under study ({\em e.g.}, droplet concentration) are defined. It is possible to reproduce all the results from the examples discussed below by following the instructions contained in \texttt{README\_howToPreparation} and \texttt{README\_howToRun} for the preparation and simulations steps, respectively.

\subsection{Output files options}\label{sec:IO}

\noindent In this paragraph, we describe the parameters that can be set in the input file \texttt{tlbfind.inp} controlling the output files properties, such as dump frequency and format type. All ASCII files including two-dimentional fields can be plotted by using the \texttt{pm3d} command of \texttt{gnuplot}. The VTK files can be read with programs like Paraview~\cite{Ahrens_2014}. Table~\ref{table:output1} reports typical values of the output parameters.\\
Independently from the output options that are set, the program will output a few files at the beginning of the preparation step, containing the information about the initial state of the various fields describing the system. These files are \texttt{init\_rho1.dat}, \texttt{init\_rho2.dat}, and \texttt{init\_temperature.dat}, including the initial values of the density fields of the two components and the temperature field, respectively.

First of all we describe the options related to the density and velocity fields of the two components. The main option is \texttt{nout density} which is an integer setting the time interval for the ASCII dumps \texttt{firstdensity.\#.dat} and \texttt{seconddensity.\#.dat}. The program also writes the files \texttt{firstdensity.dat} and \texttt{seconddensity.dat} containing the last dumped values and which are overwritten at each dump. If \texttt{nout density} is set to a value \texttt{$\leq$ 0} no dump is performed. The variable \texttt{write vtk file} (Boolean), acts as a primary switch changing the dumpfile format from ASCII to binary VTK 2.0 files. In order to assure the VTK dump one needs to toggle the (Boolean) variables \texttt{write vtk file rho1} and \texttt{write vtk file rho2} which will make the program write on disk \texttt{firstdensity.\#.vtk} and \texttt{seconddensity.\#.vtk}, respectively, with \texttt{firstdensity.vtk} and \texttt{seconddensity.vtk} containing the last dumped configurations. The variable \texttt{nout velocity} sets the time interval for the ASCII dump number of steps for the vector velocity field, \texttt{veloconf.\#.dat}, with the last configuration dumped in \texttt{veloconf.dat}. If \texttt{write vtk file} is set to 1 the binary VTK 2.0 files \texttt{veloconf.\#.vtk} are also dumped.

The variable \texttt{nout temperature} is an integer setting the time interval for the ASCII dumps \texttt{temperature.\#.dat}, with \texttt{temperature.dat} containing the last configuration. The variable \texttt{write vtk file temperature} (Boolean), toggles the dumps in VTK 2.0 binary file \texttt{temperature.\#.vtk}.\\
The variable \texttt{write energy file} (Boolean) is the main switch for writing the values of the global kinetic energy $E = \langle |{\bf u}|^2 \rangle_{x,y} / (\mbox{\texttt{nx}} \cdot \mbox{\texttt{ny}})$ that are appended every \texttt{nout energy} steps to the file \texttt{timeEnergy.dat}. The variable \texttt{nout tensor} sets the number of steps for dumping the stress tensor ASCII files \texttt{Ptot\_xy.\#.dat}, with \texttt{Ptot\_xy.dat} containing the last configuration. The variable \texttt{nout average} is an integer for the number of steps for dumping the average of the $x$-component in the $x$-direction of (i) the velocity field in \texttt{u\_av.\#.dat} files, and (ii) the stress tensor in \texttt{Pxy\_av.\#.dat} files. The last configurations are dumped in \texttt{u\_av.dat} and \texttt{Pxy\_av.dat} files, respectively. More details on the structure of the above-mentioned output files can be found in the \texttt{README\_Output} file.

Concerning the variables used for dumping the state of the system in order to allow the simulation restart, \texttt{noutconfig} is an integer setting the dumping number of steps for  binary files \texttt{conf1\_\#.in}, \texttt{conf2\_\#.in}, and \texttt{confG\_\#.in}, containing populations for the first component, second component, and temperature, respectively. Furthermore, the number of successive populations files is written in a sequence, modulo \texttt{noutconfigMod}, for fail-safe purposes. 

Some variables in \texttt{tlbfind.inp} are referred to the detection of plastic events. The integer \texttt{ncheckdelaunay} sets the number of steps at which the program checks for T1 plastic events (see Section~\ref{subsec:delaunay}). The output files are written in the sub-directories \texttt{./delaunayTriggerDir}, collecting the dumps related to plastic events, and \texttt{./delaunayNoTriggerDir}, when no plastic event occurs. Within these folders, the Delaunay binary files \texttt{delaunayNowTime\#}, \texttt{delaunayPastTime\#}, and \texttt{delaunayIsoTriggerTime\#} are dumped, containing information about the position of the centre-of-mass of all droplets, and the links connecting nearest droplets. By inspecting the functions contained in \texttt{delaunayCuda.cu}, one can find the details about the structure of those binary files. Furthermore, the trigger for the detection of plastic events also dumps a few ASCII files in the Delaunay directories that can be used for a preliminary run-time analysis of the simulation itself. In the files \texttt{arisingLinks} and \texttt{arisingLinksBoundary} the user can find data of the triangulation links that are formed after a plastic event. On the other hand, \texttt{breakingLinks} and \texttt{breakingLinksBoundary} files store data about the links vanishing after a plastic event is detected. Finally, the time sequence of the total number of droplets and the total number of links in the Delaunay triangulation are contained in \texttt{nBubblesOut} and \texttt{nLinksOut}, respectively. A detailed description of these files is contained in the \texttt{README\_Output} file.

\subsection{Preparation}\label{sec:preparation}

All simulation parameters in \texttt{tlbfind.inp} used in the preparation step are listed in Table~\ref{table:prep}. At this step, no driving force or coupling with the temperature field has to be considered. A step-by-step description of commands to run a preparation-step simulation can be found in the \texttt{README\_howToPreparation} file.
\subsubsection{Two-component fluid}\label{subsec:2fluids}
The two-dimensional system is resolved with a regular grid with size \texttt{nx} along the stream-flow direction $x$ and \texttt{ny} along the vertical $y$-direction. As discussed in Section~\ref{subsec:GPU}, TLBfind is parallelised on GPU threads via the CUDA toolkit thus, once the simulation domain is fixed, it is extremely important to declare in the input file \texttt{Which GPU} to use~\footnote{It depends on the machine used.}, the number of used threads (\texttt{nthreads}) per block, and the number of thread blocks (\texttt{nblocks}). Make sure that the product of \texttt{nthreads} with \texttt{nblocks} is equal to the whole simulation domain \texttt{nx}$\cdot$\texttt{ny}~\footnote{We remark that in the CUDA framework the size of a block of threads should be assigned a multiple of 32, which is the size of the workload ``unit" called \emph{warp}, with the optimal performances that are typically obtained setting the minimum value at 128.}.\\
\begin{figure}[t!]
\begin{center}
\includegraphics[width=.85\linewidth]{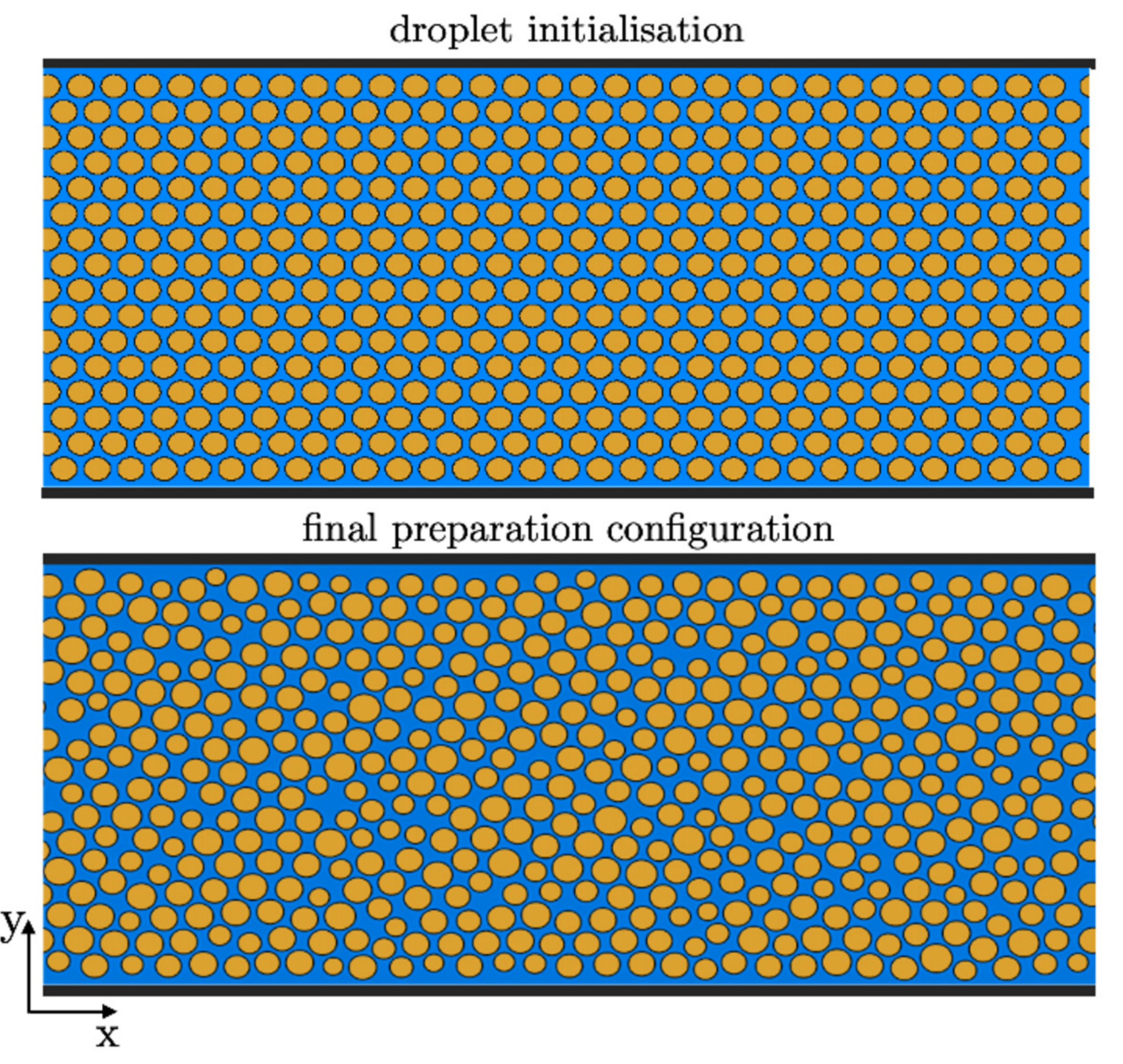}
\caption{An emulsion is prepared in an initial honeycomb structure (top panel), then it settles in a structured configuration (bottom panel). This situation is taken as the initial condition of a convection simulation experiment (see Section~\ref{sec:simulation}). At the preparation step, the temperature field is fixed to be linear between the walls, but it does not influence the fluid flow (\texttt{alphaG = 0}). For details on how to produce these plots see the \texttt{README\_howToPreparation} file. 
\label{fig:preparation}}
\end{center}
\end{figure}
\noindent Each component comprises bulk regions with initial densities assigned via the parameters \texttt{rhoMax} and \texttt{rhoMin}. The bulk densities will change during the simulation to match the values compatible with the interactions chosen. Thus, it is preferable to choose the values of \texttt{rhoMax} and \texttt{rhoMin} once the phase-separation diagram is obtained from a dedicated experiment. For an easier understanding, in Fig.~\ref{fig:G12}(a) we report the phase-separation diagram, showing the densities of each component as a function of ${\cal G}_{12}$ (for fixed competing interactions). The corresponding density values are given in~\ref{fig:G12}(b). More details on the phase-separation experiment are given in~\ref{app:G12}.\\
The bare viscosity $\eta_{0}$ in Eq.~\eqref{eq:VISCO} is fixed by the relaxation time \texttt{tau} in the LB equation~\eqref{eq:LBM}. Moreover, the strength of the phase segregation interaction in Eq.~\eqref{eq:FAB} can be tuned with \texttt{G12} and \texttt{rho0}, and the competing interactions in Eqs.~\eqref{eq:FintAttrattive} and~\eqref{eq:FintRepulsive} are handled via \texttt{G11a}, \texttt{G22a}, \texttt{G11r}, and \texttt{G22r}~\footnote{In the nomenclature of these constants, the numbers are referred to the species (1 and 2) and the $a$ and $r$ indicate the nature of the interaction (attractive or repulsive).}.
The droplets are created via the balance of phase segregation and competing interactions and can be initialised in many different ways. In the present version of TLBfind, we initialise the soft domains in an honeycomb structure (\texttt{droplet initialisation = 1}), where we establish the total number of droplets $N_{\mbox{\tiny droplets}}$ by calibrating the relative number of droplets along $x$ and $y$~\footnote{$N_{\mbox{\tiny droplets}} =$ \texttt{number of droplets x}$\cdot$ \texttt{number of droplets y}}, the droplet \texttt{diameter}, and the (\texttt{spacing}) between two consecutive droplets (see Fig.~\ref{fig:preparation}, top panel).
Then the system settles in a structured configuration, which will be the starting point of the convection simulation step (see Fig.~\ref{fig:preparation}, bottom panel).\\
In this test case, we sketch out a system confined between two walls in the $y$-direction~\footnote{It means that \texttt{periodic boundary condition along y} = 0} with \texttt{periodic boundary condition along x} as shown in Fig.~\ref{fig:preparation}~\footnote{In TLBfind no wall is implemented with the normal parallel to the $x$-direction.}. By considering flat walls, the Boolean rough-wall-flags, \texttt{roughWallUp} and \texttt{roughWallDown}, must be fixed to zero~\footnote{Note that, if \texttt{roughWallUp} and \texttt{roughWallDown} are true and \texttt{periodic boundary condition along y} = 1, simultaneously, there will be an inconsistency and the code will not work correctly.}, and the wetting properties have to be properly calibrated by changing the densities of the two species at wall with \texttt{rhoWallMax} and \texttt{rhoWallMin}, respectively.
A \texttt{flag.dat} file containing the wall (binary) density field is necessary to perform every simulation: it is the output file of the \texttt{buildObstacle} program, which in turn reads input parameters from \texttt{inputflag.inp}. The parameters used for the flat walls case are listed in Table~\ref{table:flatWall}, and they will be discussed in detail in the next Section~\footnote{We wish to remark that the wall parameters shared in \texttt{inputflag.inp} and \texttt{tlbfind.inp} files should be equal.}.\\
The above-listed information is sufficient to \texttt{start from scratch} (Boolean, true) a simulation of a static fluid for a number of \texttt{nsteps} of simulation times.

\subsubsection{Rheological characterisation}\label{subsec:rheology}
\begin{figure}[t!]
\begin{center}
\includegraphics[width=.7\linewidth]{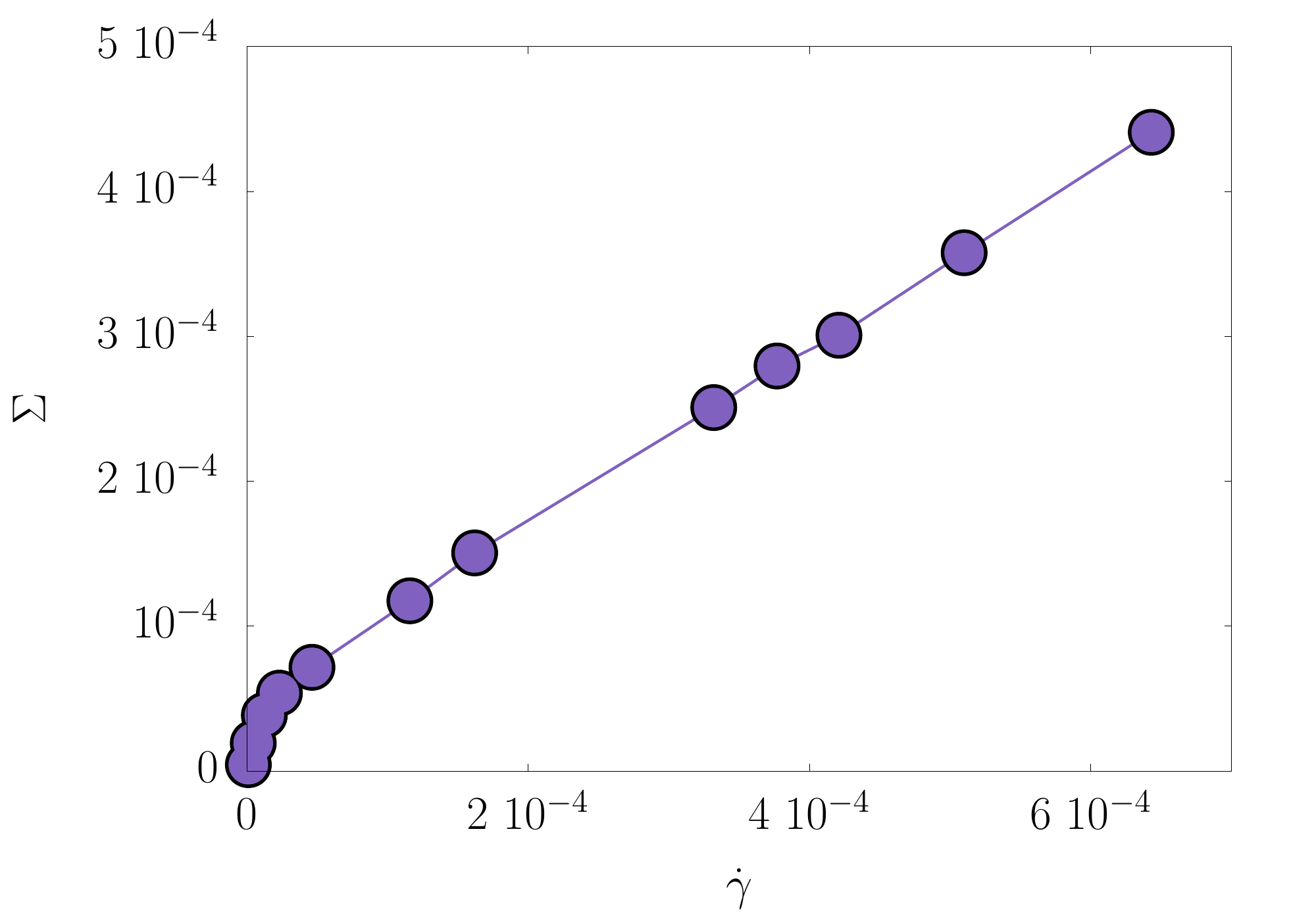}
\caption{Rheological characterisation of the emulsion shown in Fig.~\ref{fig:preparation}: the stress $\Sigma$ is reported as a function of the shear rate $\dot{\gamma}$. All dimensional quantities are given in simulation units.\label{fig:rheology}}
\end{center}
\end{figure}

Before performing a simulation of an emulsion under thermal convection, we need to learn about the system response to an applied external driving: a shear rheology experiment is of fundamental importance for this purpose. The rheological characterisation is carried out in a Couette channel~\cite{Larson,Pal96}, where a shear flow is applied through walls moving along $x$ with opposite directions, with velocities \texttt{uWallUp} and \texttt{uWallDown}. Fig.~\ref{fig:rheology} shows the flow curve for an emulsion initialised in Fig.~\ref{fig:preparation}: in this experiment, the shear rate $\dot{\gamma}$ can be calculated as
\begin{equation}
\dot{\gamma} = \dfrac{\mbox{\texttt{uWallUp}}-\mbox{\texttt{uWallDown}}}{\mbox{\texttt{ny}}},
\end{equation}
while the stress $\Sigma$ is computed by averaging in space and time the total stress, given by the sum of all components contributions~\footnote{This information is dumped in \texttt{Pxy\_av.\#.dat} files (for more details on the output files structure see the \texttt{README\_Output} file and Section~\ref{subsec:multiLBM} as a reference).}. 
%
\subsubsection{Temperature field}
One of the key point of TLBfind lies in the possibility to simulate the dynamics of the temperature field $T({\bf x},t)$ in Eq.~\eqref{eq:LBMg}, and couple it with the momentum dynamics. One can decide to activate the dynamics of $T({\bf x},t)$ by setting to \texttt{1} the Boolean parameter \texttt{THERMAL}. The thermal relaxation time \texttt{tauG} ($\tau_g$ in Eq.~\eqref{eq:LBMg}) impacts the thermal diffusivity $\kappa$ (see Eq.~\eqref{eq:K}).\\
In the setup of the Rayleigh-B\'{e}nard thermal convection, the system is confined between two walls at a different temperature: a hot bottom wall at temperature \texttt{Tdown} and a top cold one at temperature \texttt{Tup} (see Fig.~\ref{fig:RBCflat}). The temperature profile is here initialised with a linear profile, corresponding to the \texttt{temperature initialization=2}. If needed (e.g. study of situations at constant temperature) the temperature can be initialised with a constant profile (\texttt{temperature initialization=0}). 
%
\subsection{Simulation in convection}\label{sec:simulation}
All values of simulation parameters in \texttt{tlbfind.inp} that need to be changed to start a convective simulation are listed in Table~\ref{table:conv}. A step-by-step description of the commands needed to run a simulation in convection can be found in the \texttt{README\_howToRun} file.
\subsubsection{The onset of convection}\label{subsec:conv}

At this point, we are ready to perform a simulation of an emulsion in convection. The onset of convection is known to be dependent on the magnitude of the buoyancy term in Eq.~\eqref{eq:alphaG}: at fixed walls temperatures, system size, and fluid properties, \texttt{alphaG} is the order parameter that triggers the transition from a conductive to a convective state.
Note that, a finite perturbation is necessary to destabilise the conductive state; this \texttt{initial velocity perturbation} depends on the emulsion rheology~\cite{Zhang06}: the more the system is non-Newtonian, the greater the required perturbation is.\\
\\
With the aim to restart a simulation from the last preparation configuration (i.e., situation in bottom Panel of Fig.~\ref{fig:preparation}), TLBfind requires four files: the three final populations files of the components 1,2 and temperature from the preparation step (i.e., \texttt{conf1.in\_0}, \texttt{conf2.in\_0}, and \texttt{confG.in\_0}, respectively); the \texttt{dumpcount.in} file, which counts the enumeration of time steps and dumping number of output files~\footnote{The \texttt{dumpcount.in} file must be manually created.}. At $t = 0$ it is equal to a series of seven zeros \texttt{0 0 0 0 0 0 0}; it is very useful especially in the case of a restart of a finished or suspended simulations~\footnote{For details on how to restart a finished or suspended simulation see the \texttt{README\_howToRestart} file.}.
As highlighted in Table~\ref{table:conv}, at this step \texttt{start from scratch} has to be switched off while \texttt{post preparation temp} has to be switched on~\footnote{It is helpful to distinguish the case of simulation after preparation from a simple restart: in the latter case no initial velocity perturbation has to be applied, thus \texttt{post preparation temp} is false.}.\\
%
\begin{figure}[t!]
\begin{center}
\includegraphics[width=.95\linewidth]{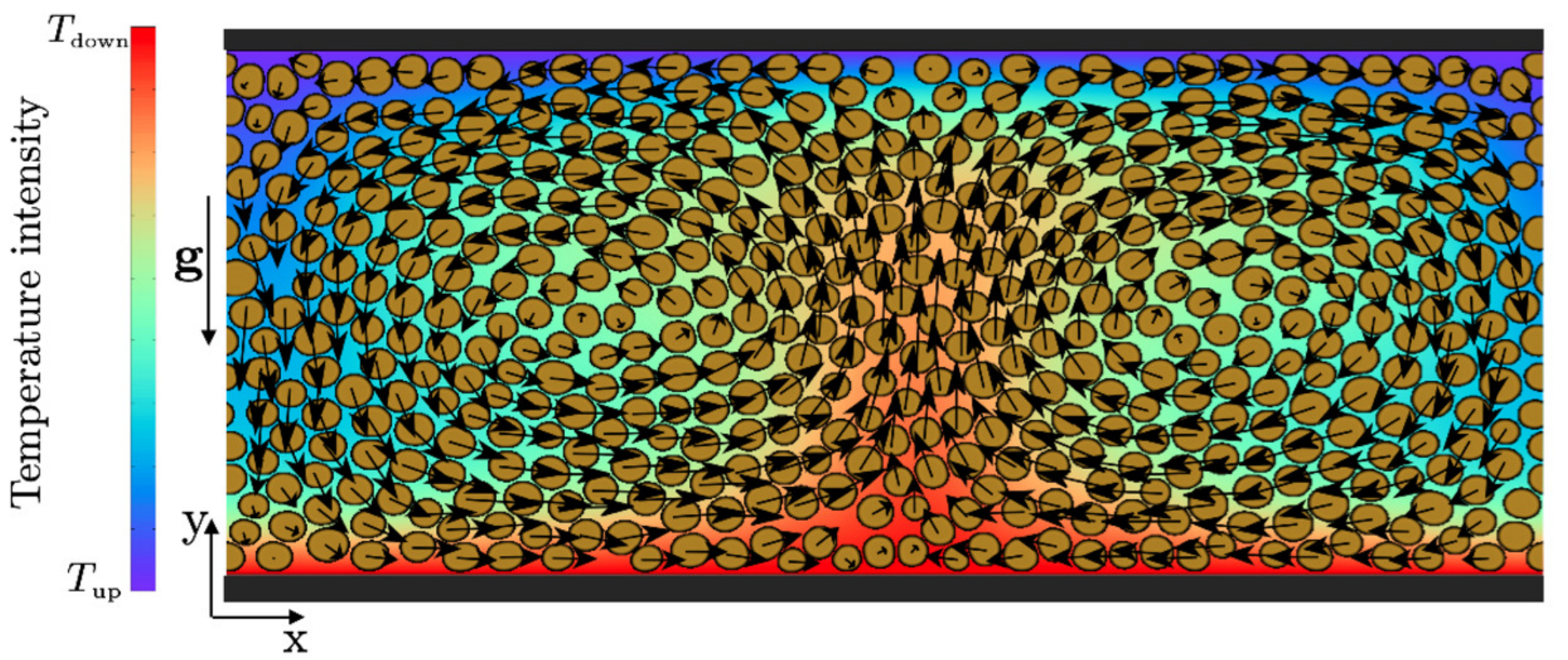}
\caption{Rayleigh-B{\'e}nard convection experiment: the convection simulation step. Black arrows refer to droplet displacements: the typical convective rolls are visible. For details on how to produce this plot see the \texttt{README\_howToRun} file.\label{fig:RBCflat}}
\end{center}
\end{figure}
%
Fig.~\ref{fig:RBCflat} shows the emulsion prepared in Fig.~\ref{fig:preparation} when subjected to a sufficiently large buoyancy force: the convective state, with the characteristic circular convective rolls, is highlighted thanks to the overlap to the density map of Lagrangian droplet displacement vector fields $\mbox{\bf d}(\mathbf{X}_i(t),t)$ (black arrows), computed for all droplet centre-of-mass $\mathbf{X}_i(t)$ ($i=1...N_{\mbox{\tiny droplets}}$). Both centre-of-mass positions and displacement of all droplets at any time can be computed as part of the Delaunay analysis (see Section~\ref{subsec:delaunay}). The threshold used in order to identify the droplets region can be tuned via the \texttt{bubbleThreshold} (i.e., $B_{th}$) parameter: it defines the floating-point value used as a threshold on the first density field to detect the high-density compact regions constituting the droplets. Then, \texttt{nmindelaunay} sets the starting time for the Delaunay analysis. This can be useful for discarding the initial transient of a flow from the analysis. From the knowledge of centre-of-mass positions, the computation of droplet displacement is an immediate consequence. As mentioned in Section~\ref{sec:IO}, Delaunay binary files can be analysed in order to extract droplets information by compiling the source codes of the following two programs:
\begin{enumerate}
    \item \texttt{deltaAnalysis}, which executes the analysis for the calculation of the displacement of the centre of mass of the droplets yielding further information as well; the main output is contained in the ASCII file \texttt{DeltaField}, see the \texttt{README\_Analysis} file for further details;
    \item \texttt{dropletStats}, which appends in the ASCII files \texttt{dropletsHostNow} and \texttt{dropletsHostPast} some relevant information about the droplets, such as centre-of-mass coordinates and size (see the \texttt{README\_Analysis} file for details).
\end{enumerate}


\subsubsection{Heat transfer analysis}\label{subsec:heat}

\begin{figure}[t!]
\begin{center}
\begin{tabular}{c}
\includegraphics[width=.97\linewidth]{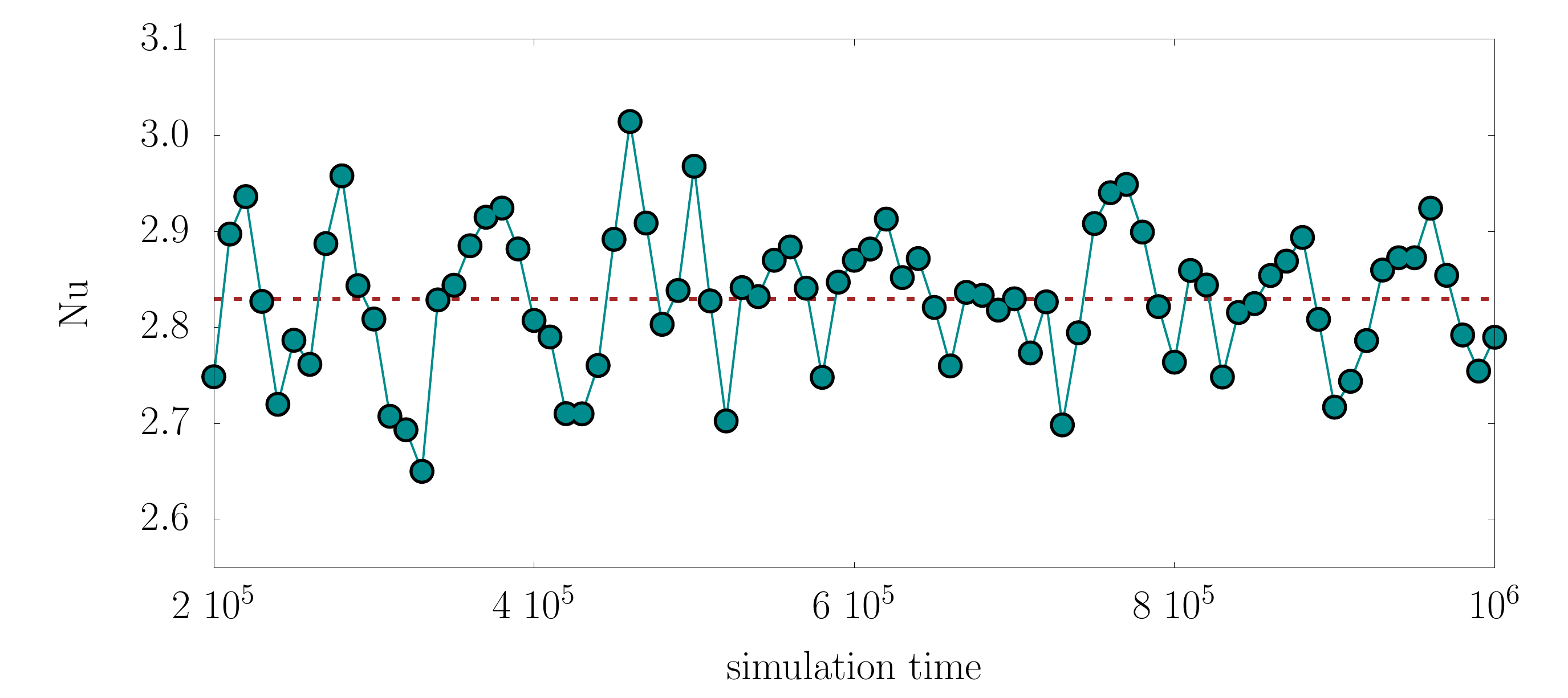}\\
\hspace{0.5cm} \small (a) \\
\includegraphics[width=.9\linewidth]{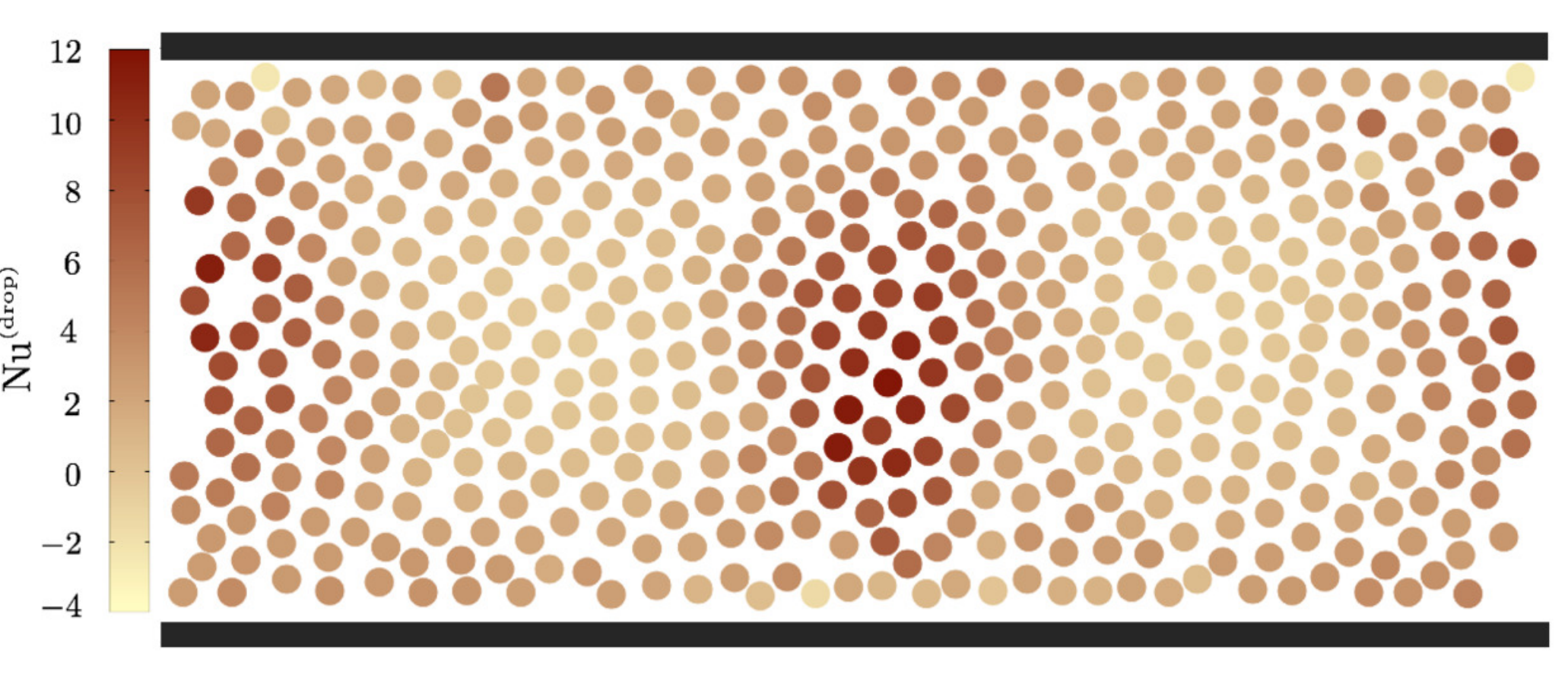}\\
 \hspace{0.5cm} \small (b)
  \end{tabular}
\caption{Panel (a): Time behaviour of the Nusselt number defined in Eq.~\eqref{eq:NuTime} (expressed in simulation units) for emulsion of Fig.~\ref{fig:RBCflat} in the statistically steady state. The average Nusselt number is indicated with the dashed line. Panel (b): a snapshot of the droplet Nusselt number defined in Eq.~\eqref{eq:NuDrops}. For further details on how producing these plots see the \texttt{README\_howToRun} file.\label{fig:nusseltFlat}}
\end{center}
\end{figure}
The heat-flux properties of the system can be studied in TLBfind. A measure of the average heat flux can be performed in terms of the Nusselt number (Nu), a dimensionless observable defined as~\cite{Shraiman90,Ahlers09,Verzicco10,Chilla12}
\begin{equation}\label{eq:NuTime}
\mbox{Nu}(t) = \frac{\langle u_y (x,y,t) T (x,y,t)\rangle_{x,y} - \kappa \langle \partial_y T (x,y,t) \rangle_{x,y}}{\kappa \frac{\Delta T}{\mbox{\texttt{ny}}}},
\end{equation}
where $\langle (\dots) \rangle_{x,y}$ denotes a space average, and $\Delta T = \mbox{\texttt{Tdown}}-\mbox{\texttt{Tup}}$. $\mbox{Nu}$ gives an estimate of the balance between convective and conductive transport at macroscopic scales. The program \texttt{nusseltNumber} allows computing Nu as a function of time (see the \texttt{README\_Analysis} file for details), as Fig.~\ref{fig:nusseltFlat}(a) shows for the emulsion displayed in Fig.~\ref{fig:RBCflat}. Because of the finite-size effects at mesoscales~\cite{PelusiSM21}, in some cases there may be the need to deal with the heat flux contributions of the single droplets. For this reason, a Nusselt number associated with each single droplet can be defined as~\cite{PelusiSM21}: 
\begin{equation}\label{eq:NuDrops}
\mbox{Nu}^{(\mbox{\tiny drop})}_{i}(t) = \frac{u^{(i)}_y(t)  T^{(i)}(t) - \kappa (\partial_yT)^{(i)}(t)}{\kappa \frac{\Delta T}{\mbox{\texttt{ny}}}},
\end{equation}
where $u^{(i)}_y(t)=u_y(\mathbf{X}_i(t),t)$, $T^{(i)}(t)=T(\mathbf{X}_i(t),t)$ and $(\partial_yT)^{(i)}(t)=\partial_y T(\mathbf{X}_i(t),t)$ are the fluid velocity, temperature, and temperature gradient evaluated for the $i$-th droplet, respectively.

The program \texttt{nusseltNumberDroplet} computes Eq.~\eqref{eq:NuDrops} for all droplets at any simulation time step by reading velocity and temperature fields and combining them with the centres-of-mass positions extracted from Delaunay analysis (see the \texttt{README\_Analysis} file for further details).

\section{Test case with rough walls}\label{sec:testRough}
\begin{figure}[t!]
\begin{center}
\begin{tabular}{c c}
\includegraphics[width=.485\linewidth]{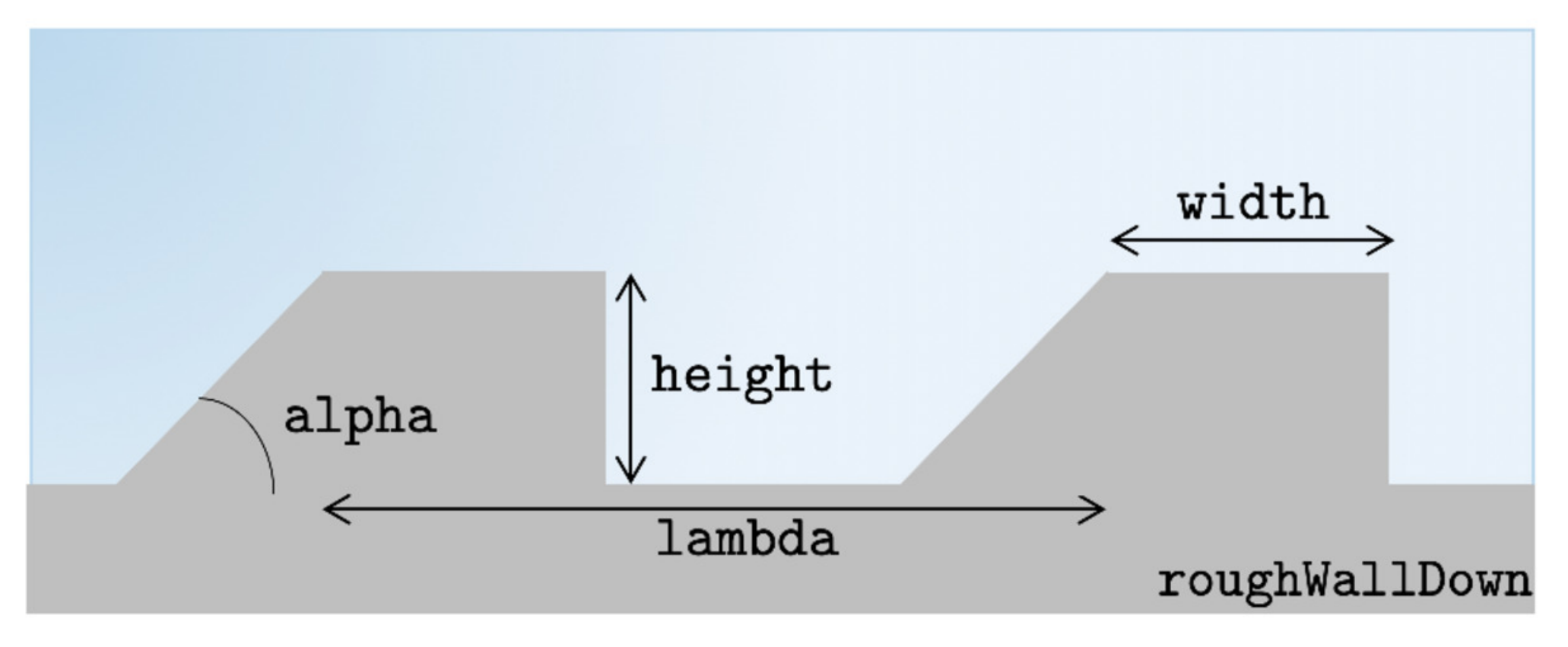} & \includegraphics[width=.485\linewidth]{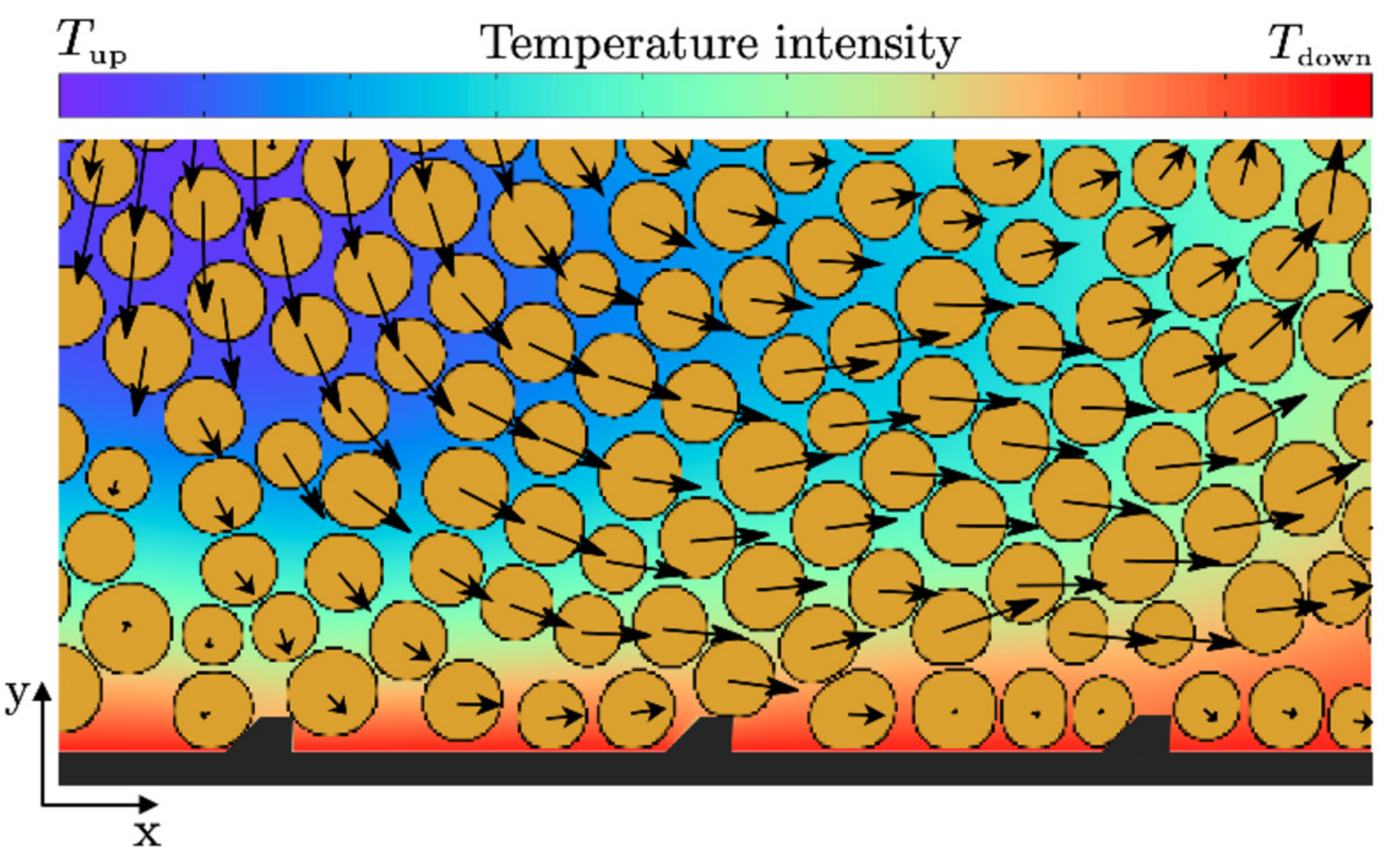} \\
\small (a) & \small (b)\\
\end{tabular}
\caption{Panel (a): sketch of a roughness on the bottom wall (\texttt{roughWallDown} = 1): the obstacles have asymmetric trapezoidal-shape with inclination given by the angle \texttt{alpha}, specific \texttt{height}, \texttt{width}, and repeated every \texttt{lambda} lattice nodes. Panel (b): a snapshot of Rayleigh-B{\'e}nard cell in the proximity to the wall during a convection experiment. Black arrows indicate the droplet displacement field.\label{fig:roughness}}
\end{center}
\end{figure}

In the previous section, we have discussed a test case of Rayleigh-B{\'e}nard numerical experiment in the presence of flat walls.Further complexity may be introduced by the presence of roughness along the walls. Here we report the case of rough walls with an asymmetric trapezoidal shape, as shown in Fig.~\ref{fig:roughness}(a). The \texttt{height}, the \texttt{width}, the periodicity (via \texttt{lambda}), and the angle (via \texttt{alpha}) can be varied. Note that the rough-wall-flags \texttt{roughWallUp} and \texttt{roughWallDown} have to be switched on in order to design the roughness at the walls.
Table~\ref{table:roughWall} lists roughness parameters in the \texttt{inputflag.dat} file used in the present test case. We build here a very general roughness shape by accessing all input parameters involved. As a result of this code, a symmetric trapezoidal-shape roughness was used in~\cite{PelusiEPL19}, where simulations with a symmetric roughness (\texttt{asymmetric = 0}) built only on the bottom wall (\texttt{roughWallUp = 0}) were performed.
Moreover, in a very similar context~\cite{Derzsi17,Derzsi18}, the dynamics of emulsions in a channel with a simple rectangular-shape roughness (\texttt{alpha = 0.0}) was analysed in detail.\\
In order to perform simulation of Rayleigh-B{\'e}nard convection, the steps to follow are the same as shown for the flat-walls case: after the system preparation step, we run the simulation with the emulsion in convection. For both these steps, we modified the input files as well. Input parameters changed with respect to the flat case are shown in Table~\ref{table:roughConv}. A snapshot of an emulsion in convection with these boundary conditions is shown in Fig.~\ref{fig:roughness}(b).

\section{Conclusions}\label{sec:conclusions}
We have presented TLBfind, a high-performance CUDA code for the simulation of thermal flows of finite-size droplet suspensions with non-trivial boundary conditions. TLBfind implements a multi-component Lattice Boltzmann (LB) method where the non-ideal interactions are tuned in such a way to promote inhibition of coalescence of adjacent droplets. With the LB method, it is possible to simulate advection and diffusion of a temperature field coupled to the droplets dynamics, and easily implement boundary conditions with complex geometry. We demonstrated various code features by means of two examples, corresponding to the case with flat and rough walls. We expect to work in the future on the extension of the present code to three dimensions. A multi-GPU implementation is also part of our future plans.
\section{Acknowledgements}
We thank Roberto Benzi, Andrea Scagliarini and Sauro Succi. FP thanks Fabio Bonaccorso for technical support in accessing and operating GPUs at the University of Rome Tor Vergata. ML acknowledges the support of the National Science Foundation of China Grant~12050410244.
\appendix

\section{Input parameters lists}\label{}

\begin{table}[h!]
\begin{center}
\begin{tabular}{|p{5.cm} p{1.cm}| p{4.5cm} p{1.5cm}|}
\hline
\rowcolor{GreenYellow} \multicolumn{4}{|c|}{Output files parameters} \\
\hline
\texttt{nout density} & 10000 & \texttt{write energy file} & 1\\
\texttt{write vtk file} & 0 & \texttt{nout energy} & 1000\\
\texttt{write vtk file rho1} & 1 & \texttt{nout tensor} & 10000000\\
\texttt{write vtk file rho2} & 0 & \texttt{nout average} & 1000\\
\texttt{nout velocity} & 1000 & \texttt{noutconfig} & 1000\\
\texttt{nout temperature} & 1000 & \texttt{noutconfigMod} & 2\\
\texttt{write vtk file temperature} & 0 & \texttt{ncheckdelaunay} & 1000\\
\hline
\end{tabular}
\caption{Output files parameters in \texttt{tlbfind.inp}. All ASCII files of the various two-dimensional fields can be plotted with \texttt{pm3d} on \texttt{gnuplot}. All VTK files are VTK 2.0 binary files.}\label{table:output1}
\end{center}
\end{table}
\begin{table}[h!]
\begin{center}
\begin{tabular}{|p{5.2cm} p{.8cm}| p{5.2cm} p{.8cm}|}
 \hline
\rowcolor{GreenYellow} \multicolumn{4}{|c|}{Input preparation parameters} \\
\hline
\texttt{nx} & 947 & \texttt{number of droplet x} & 31\\
\texttt{ny} & 431 & \texttt{number of droplet y} & 16\\
\texttt{rhoMax} & 1.4 & \texttt{diameter} & 23.92\\
\texttt{rhoMin} & 0.1 & \texttt{spacing} & 6.4\\
\texttt{tau} & 1.0 & \texttt{WD} & 0.6\\
\texttt{bubbleThreshold} & 0.8 & \texttt{threhold\_WD} & 0.4\\
\texttt{rho0} & 0.83 & \texttt{THERMAL} & 1\\
\texttt{G12} & 0.405 & \texttt{temperature initialisation} & 2 \\
\texttt{G11a} & -9.0 & \texttt{Tup} & -0.5\\
\texttt{G22a} & -8.0 & \texttt{Tdown} & 0.5\\
\texttt{G11r} & -8.1 & \texttt{tauG} & 1.0\\
\texttt{G22a} & -7.1 & \texttt{alphaG} & 0.0\\
\texttt{rhoWallMax} & 0.612 & \texttt{periodic boundary condition along x} & 1\\
\texttt{rhoWallMin} & 0.612 & \texttt{periodic boundary condition along y} & 0\\
\texttt{start from scratch} & 1 & \texttt{nsteps} & $10^{5}$\\
\texttt{droplet initialisation} & 1 & \texttt{initial velocity perturbation} & 0.0\\
\texttt{delaunayDebug} & 0 & & \\
\hline
\end{tabular}
\caption{Preparation parameters in \texttt{tlbfind.inp}. \texttt{WD} and \texttt{threshold\_WD} are used to introduce randomization in the initial condition; the used values produce the initial condition displayed in Fig.~\ref{fig:preparation}.}\label{table:prep}
\end{center}
\end{table}
\begin{table}[h!]
\begin{center}
\begin{tabular}{|p{5.2cm} p{.8cm}| p{4.5cm} p{1.6cm}|}
 \hline
\rowcolor{GreenYellow} \multicolumn{4}{|c|}{Input flat wall parameters} \\
\hline
\texttt{roughWallUp} & 0 & \texttt{width} & 0\\
\texttt{roughWallDown} & 0 & \texttt{alpha} & 0.0\\
\texttt{lambda} & 0 & \texttt{asymmetric} & 0\\
\texttt{height} & 0 & &\\
\hline
\end{tabular}
\caption{Input parameters in \texttt{inputflag.inp} for a flat wall.}\label{table:flatWall}
\end{center}
\end{table}
\begin{table}[h!]
\begin{center}
\begin{tabular}{|p{5.2cm} p{.8cm}| p{4.5cm} p{1.6cm}|}
 \hline
\rowcolor{GreenYellow} \multicolumn{4}{|c|}{Input simulation parameters} \\
\hline
\texttt{start from scratch} & 0 & \texttt{alphaG} & $1.24 \ 10^{-5}$\\
\texttt{nsteps} & $10^{6}$ & \texttt{initial velocity perturbation} & $10^{-4}$\\
\texttt{post preparation temp} & 1 & & \\
\hline
\end{tabular}
\caption{Input parameters varied in \texttt{tlbfind.inp} to start a convective simulation after the preparation step.}\label{table:conv}
\end{center}
\end{table}
\begin{table}[h!]
\begin{center}
\begin{tabular}{|p{5.2cm} p{.8cm}| p{4.5cm} p{1.6cm}|}
 \hline
\rowcolor{GreenYellow} \multicolumn{4}{|c|}{Input rough wall parameters} \\
\hline
\texttt{roughWallUp} & 1 & \texttt{width} & 10\\
\texttt{roughWallDown} & 1 & \texttt{alpha} & 0.25\\
\texttt{lambda} & 150 & \texttt{asymmetric} & 1\\
\texttt{height} & 12 & &\\
\hline
\end{tabular}
\caption{Input parameters in \texttt{inputflag.inp} for a rough wall as in Fig.~\ref{fig:roughness}.}\label{table:roughWall}
\end{center}
\end{table}
\begin{table}[h!]
\begin{center}
\begin{tabular}{|p{4.5cm} p{1.6cm}| p{5.2cm} p{.8cm}|}
 \hline
\rowcolor{GreenYellow} \multicolumn{4}{|c|}{Input simulation parameters} \\
\hline
\texttt{roughWallUp} & 1 & \texttt{rhoWallMax} & 6.0\\
\texttt{roughWallDown} & 1 & \texttt{rhoWallMin} & 0.1\\
\hline
\end{tabular}
\caption{Input parameters varied (compared to the flat wall case) in \texttt{tlbfind.inp} to run the simulation with rough walls discussed in the text (see Tables~\ref{table:prep} and~\ref{table:conv}).}\label{table:roughConv}
\end{center}
\end{table}

\section{Dependence of densities $\rho_1$ and $\rho_2$ on the coupling parameter ${\cal G}_{12}$}\label{app:G12}

\begin{figure}[ht!]
\centering
\begin{tabular}{c}
\includegraphics[width=.85\linewidth]{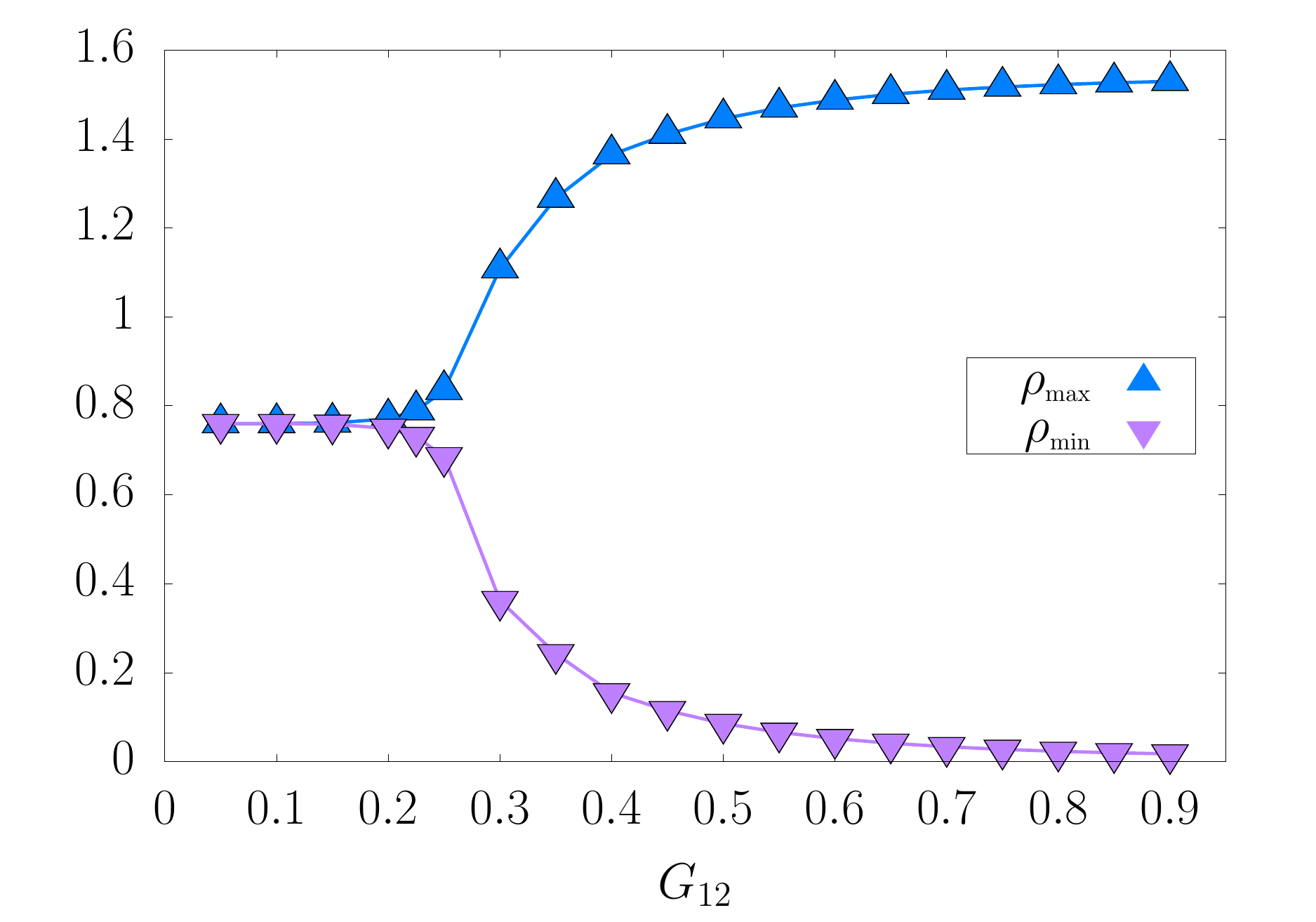} \\
\small (a) \\
\vspace{0.1cm}\\
\begin{tabular}{|| c c c | c c c ||}
\hline
${\cal G}_{12}$ & $\rho_{\mbox{\tiny max}}$ & $\rho_{\mbox{\tiny min}}$ & ${\cal G}_{12}$ & $\rho_{\mbox{\tiny max}}$ & $\rho_{\mbox{\tiny min}}$\\
\hline
0.05  &  0.7601748 & 0.7601377 & 0.50  &  1.445688 & 0.08590994\\
0.10   &  0.7603156 & 0.7599969 & 0.55  &  1.470143 & 0.06602583\\
0.15  &  0.7614925 & 0.7588201 & 0.60   &  1.487723 & 0.05179497\\
0.20    & 0.7708472 & 0.7494691 & 0.65  &  1.500535 & 0.04135837\\
0.25  &  0.8346988 & 0.6855015 & 0.70   &  1.509980 & 0.03360042\\
0.30  &   1.108622 & 0.3624054 & 0.75  &  1.517030 & 0.02778616\\
0.35 &   1.267322 & 0.2425093 & 0.80  &  1.522366 & 0.02340400\\
0.40   &  1.364685 & 0.1549724 & 0.85  &  1.526470 & 0.02008613\\
0.45  &  1.410851 & 0.1149639 & 0.90   &  1.529679 & 0.01756290\\
\hline
\end{tabular}\\
\small (b) \\
\end{tabular}
\caption{Panel (a): phase-separation diagram for a system with initial densities \texttt{rhoMax}=1.4 and \texttt{rhoMin}=0.1 and fixed competing interactions given in Table~\ref{table:G12}. Panel (b) shows values of densities as a function of ${\cal G}_{12}$ producing the upper plot.\label{fig:G12}}
\end{figure}

\noindent As discussed in Section~\ref{subsec:2fluids}, initial bulk densities are assigned via the parameters \texttt{rhoMax} and \texttt{rhoMin}. A dedicated phase-separation experiment is needed in order to set these values in such a way that they will not change sensibly during the simulation. Here we provide an experiment example: we performed some dedicated simulations with parameters in Table~\ref{table:G12}, and we measured the maximum and minimum density values from the last configuration of the first component (see Fig.~\ref{fig:G12}(a)) by varying only the value of \texttt{G12}~\footnote{The same is valid if the measure is done with the second component, but with an inversion of the maximum and minimum values.}. In this kind of experiment, as well as for the test cases presented in the text, the preparation-step features must be followed, as explained in Section~\ref{sec:preparation}, with \texttt{droplet initialization = 3}. With this initialisation, we prepared a simple system of two layers: one half of the domain is occupied with the majority of the first component, and the second half with the second component. Then, we apply no force, but we wait a sufficient number of time steps to observe a stabilised phase separation, and we look at the last density configuration of the first component. The resulting phase-separation diagram is shown in Fig.~\ref{fig:G12}(a), and the corresponding measured values of bulk densities are listed in Table~\ref{fig:G12}(b). Note that, in this specific case, for values of ${\cal G}_{12}$ below $\approx$ 0.25, the minimum density value coincides with the maximum one: in such conditions, there is no phase separation and the two components mix.

\begin{table}[h!]
\begin{center}
\begin{tabular}{|p{5.2cm} p{.7cm}| p{5.2cm} p{.9cm}|}
 \hline
\rowcolor{GreenYellow} \multicolumn{4}{|c|}{Phase-segregation experiment parameters} \\
\hline
\texttt{nx} & 128 & \texttt{THERMAL} & 0\\
\texttt{ny} & 2 & \texttt{start from scratch} & 1 \\
\texttt{rhoMax} & 1.4 &  \texttt{nsteps} & $5 \ 10^{4}$\\
\texttt{rhoMin} & 0.1 & \texttt{periodic boundary condition along x} & 1\\
\texttt{tau} & 1.0 &  \texttt{periodic boundary condition along y} & 1\\
\texttt{rho0} & 0.83 & \texttt{droplet initialisation} & 3 \\
\texttt{G12} & 0.405 & \texttt{nout density} & 10000 \\
\texttt{G11a} & -9.0 & \texttt{write vtk file} & 0\\
\texttt{G22a} & -8.0 & \texttt{write vtk rho1} & 1\\
\texttt{G11r} & -8.1 & \texttt{noutconfig} & 10000\\
\texttt{G22a} & -7.1 & \texttt{noutconfigMod} & 2\\
\hline
\end{tabular}
\caption{Phase-segregation experiment parameters. Regarding the output files and preparation parameters present in Tables~\ref{table:output1} and~\ref{table:prep}, and not listed here, they can be all switched off because not necessary.}\label{table:G12}
\end{center}
\end{table}

\section{Output Files}

\noindent Here we schematically report the list of the output files generated by the simulation program itself (not the analysis programs) along with a brief description and the associated flags in the input file

\begin{table}[h!]
\begin{center}
\begin{tabular}{|p{6cm}|p{6cm}|}
 \hline
\rowcolor{GreenYellow} \multicolumn{2}{|c|}{Simulation output files} \\
\hline
\texttt{init\_rho1.dat} \newline
\texttt{init\_rho2.dat} \newline
\texttt{init\_temperature.dat} & Initial conditions of the components densities and the temperature fields \\
\hline
\hline
\texttt{firstdensity.\#.dat} \newline \texttt{seconddensity.\#.dat} & Components density fields at the \texttt{\#} step. Input file option: \texttt{nout density} $> 0$ \newline \\
\texttt{firstdensity.dat} \newline \texttt{seconddensity.dat} & Components density fields at the last iteration \newline \\
\texttt{firstdensity.\#.vtk} & Input file options: \texttt{write vtk file} and \texttt{write vtk file rho1} set to 1 \newline\\
\texttt{seconddensity.\#.vtk} & Input file options: \texttt{write vtk file} and \texttt{write vtk file rho2} set to 1 \newline\\
\texttt{temperature.\#.dat} & Temperature field at the \texttt{\#} step: input option \texttt{nout temperature} $>0$ \newline \\
\texttt{temperature.dat} & same as above with last iteration data \newline \\
\texttt{temperature.\#.vtk} & Input file options: \texttt{nout temperature} $>0$ and \texttt{write vtk file temperature} set to 1\\
\hline
\end{tabular}
\caption{On the left, list of simulation output files for densities and temperature fields. On the right, the relevant parameters of the input file \texttt{tlbfind.inp}.\label{table:output_1}}
\end{center}
\end{table}

\begin{table}[h!]
\begin{center}
\begin{tabular}{|p{6cm}|p{6cm}|}
 \hline
\rowcolor{GreenYellow} \multicolumn{2}{|c|}{Simulation output files} \\
\hline
\texttt{veloconf.\#.dat} & Velocity field at the \texttt{\#} step. Input file option \texttt{nout velocity}$>0$ \newline \\
\texttt{veloconf.dat} & Velocity field at the last iteration \newline \\
\texttt{veloconf.\#.vtk} & Input file options: \texttt{nout velocity} $>0$ and \texttt{write vtk file} set to 1 \newline\\
\texttt{u\_av.\#.dat} & x-wise x-component average of the velocity field. Input file options: \texttt{nout velocity} $>0$, \texttt{nout average} $>0$ and \texttt{write vtk file} set to 0 \newline \\
\texttt{u\_av.dat} & same as above with last iteration data\\
\hline
\hline
\texttt{Ptot\_xy.\#.dat} & Stress tensor field at the \texttt{\#} step. Input file option: \texttt{nout tensor} $>0$ \newline\\
\texttt{Ptot\_xy.dat} & same as above with last iteration data \newline\\
\texttt{Pxy\_av.\#.dat} & x-wise average of the stress tensor field at the \texttt{\#} step. Input file options: \texttt{nout tensor} $>0$ and \texttt{nout average} $>0$\\
\texttt{Pxy\_av.dat} & same as above with last iteration data\\
\hline
\end{tabular}
\caption{On the left, list of simulation output files for the velocity and stress tensor fields. On the right, the relevant parameters of the input file \texttt{tlbfind.inp}.}\label{table:output_2}
\end{center}
\end{table}

\begin{table}[h!]
\begin{center}
\begin{tabular}{|p{6cm}|p{6cm}|}
 \hline
\rowcolor{GreenYellow} \multicolumn{2}{|c|}{Simulation output files} \\
\hline

\texttt{timeEnergy.dat} & Input file option: \texttt{write energy file} set to 1 and \texttt{nout energy} $>0$\\
\texttt{conf1\_\#.in} & First component populations field dumped as a binary file. Input file options: \texttt{noutconfig} $>0$, \texttt{noutconfigMod} $>1$\newline \\
\texttt{conf2\_\#.in} & Second component populations field dumped as a binary file. Input file options: \texttt{noutconfig} $>0$, \texttt{noutconfigMod} $>1$ \newline\\
\texttt{confG\_\#.in} & Temperature field populations dumped as a binary file. Input file oprions: \texttt{noutconfig} $>0$, \texttt{noutconfigMod} $>1$\\
\hline
\end{tabular}
\caption{On the left, list of simulation output files for the system energy as a function of time and the binary dumps for the populations. On the right, the relevant parameters of the input file \texttt{tlbfind.inp}.}\label{table:output_3}
\end{center}
\end{table}

\begin{table}[h!]
\begin{center}
\begin{tabular}{|p{6cm}|p{6cm}|}
 \hline
\rowcolor{GreenYellow} \multicolumn{2}{|c|}{Simulation Delaunay output files} \\
\hline
\texttt{./delaunayTriggerDir} & Directory created if \texttt{ncheckdelaunay}$>0$ containing a plethora of files which are described below \newline\\

\texttt{arisingLinks} & File containing triangulation links created after a plastic event \\

\texttt{arisingLinksBoundary} & File containing links created after a plastic event one the boundary of the triangulation \newline \\

\texttt{breakingLinks} & File containing triangulation links vanishing after a plastic event\\

\texttt{breakingLinksBoundary} & File containing links vanishing after a plastic event on the boundary of the triangulation \newline \\

\texttt{nBubblesOut} & Time sequence of the number of droplets. It can be used to monitor the coalescence events \newline\\

\texttt{nLinksOut} & Time sequence of the number of links in the Delaunay triangulation \newline\\

\texttt{delaunayNowTime\#}\newline \texttt{delaunayPastTime\#}\newline \texttt{delaunayIsoTriggerTime\#} & Binary files needed for the analysis of plastic events using the programs that can be compiled from the source files \texttt{deltaAnalysis.cu} and \texttt{dropletStats.cu} \newline \\

\texttt{./delaunayNoTriggerDir} & Directory containing only the binary files described above related to the frames for which no plastic event is detected \\
\hline
\end{tabular}
\caption{On the left, list of simulation directories and output files for data related to the Delaunay triangulation and the detection of plastic events. On the right, the relevant parameters of the input file.}\label{table:output_4}
\end{center}
\end{table}



\bibliographystyle{elsarticle-num}
\bibliography{francesca.bib}

\begin{thebibliography}{10}
\expandafter\ifx\csname url\endcsname\relax
  \def\url#1{\texttt{#1}}\fi
\expandafter\ifx\csname urlprefix\endcsname\relax\def\urlprefix{URL }\fi
\expandafter\ifx\csname href\endcsname\relax
  \def\href#1#2{#2} \def\path#1{#1}\fi

\bibitem{Hetsroni82}
G.~Hetsroni, Handbook of multiphase systems, McGraw-Hill Book Co., New York,
  NY, 1982.

\bibitem{Gallegos99}
C.~Gallegos, J.~Franco, Rheology of food, cosmetics and pharmaceuticals,
  Current opinion in colloid \& interface science 4~(4) (1999) 288--293.

\bibitem{Khanetal11}
B.~A. Khan, N.~Akhtar, H.~M.~S. Khan, K.~Waseem, T.~Mahmood, A.~Rasul,
  M.~Iqbal, H.~Khan, Basics of pharmaceutical emulsions: A review, African
  Journal of Pharmacy and Pharmacology 5~(25) (2011) 2715--2725.

\bibitem{Shao15}
J.~Shao, J.~Darkwa, G.~Kokogiannakis, Review of phase change emulsions (pcmes)
  and their applications in hvac systems, Energy and buildings 94 (2015)
  200--217.

\bibitem{McClements15}
D.~J. McClements, Food emulsions: principles, practices, and techniques, CRC
  press, 2015.

\bibitem{Yukuyama16}
M.~Yukuyama, D.~Ghisleni, T.~Pinto, N.~Bou-Chacra, Nanoemulsion: process
  selection and application in cosmetics--a review, International journal of
  cosmetic science 38~(1) (2016) 13--24.

\bibitem{Wang2019}
F.~Wang, W.~Lin, Z.~Ling, X.~Fang, A comprehensive review on phase change
  material emulsions: Fabrication, characteristics, and heat transfer
  performance, Solar Energy Materials and Solar Cells 191 (2019) 218--234.

\bibitem{PrincenKiss89}
H.~Princen, A.~Kiss,
  \href{http://www.sciencedirect.com/science/article/pii/0021979789903962}{Rheology
  of foams and highly concentrated emulsions: Iv. an experimental study of the
  shear viscosity and yield stress of concentrated emulsions}, J. Colloid
  Interface Sci. 128~(1) (1989) 176 -- 187.
\newblock \href {https://doi.org/https://doi.org/10.1016/0021-9797(89)90396-2}
  {\path{doi:https://doi.org/10.1016/0021-9797(89)90396-2}}.
\newline\urlprefix\url{http://www.sciencedirect.com/science/article/pii/0021979789903962}

\bibitem{Schramm92}
L.~Schramm, Emulsions, American Chemical Society (ACS), 1992.

\bibitem{Barnes94}
H.~A. Barnes, Rheology of emulsions—a review, Colloids and Surfaces A:
  Physicochemical and Engineering Aspects 91 (1994) 89--95.

\bibitem{Coussot05}
P.~Coussot, Rheometry of Pastes, Suspensions, and Granular Materials,
  Wiley-Interscience, 2005.

\bibitem{BarratReview17}
J.-L. Barrat, Elasticity and plasticity of disordered systems, a statistical
  physics perspective, Physica A: Statistical Mechanics and its Applications
  504 (2017) 20--30.
\newblock \href {https://doi.org/10.1016/j.physa.2017.11.146}
  {\path{doi:10.1016/j.physa.2017.11.146}}.

\bibitem{Stein92}
D.~J. Stein, F.~J. Spera, Rheology and microstructure of magmatic emulsions:
  theory and experiments, Journal of Volcanology and Geothermal Research
  49~(1-2) (1992) 157--174.

\bibitem{Zhou19}
Y.~Zhou, D.~Yin, W.~Chen, B.~Liu, X.~Zhang, A comprehensive review of emulsion
  and its field application for enhanced oil recovery, Energy Science \&
  Engineering 7~(4) (2019) 1046--1058.

\bibitem{Egolf05}
P.~W. Egolf, M.~Kauffeld, From physical properties of ice slurries to
  industrial ice slurry applications, International journal of refrigeration
  28~(1) (2005) 4--12.

\bibitem{Pal2000}
R.~Pal, Shear viscosity behavior of emulsions of two immiscible liquids,
  Journal of colloid and interface science 225~(2) (2000) 359--366.

\bibitem{Larson}
R.~G. Larson, The Structure and Rheology of Complex Fluids, Oxford University
  Press, 1999.

\bibitem{Bonn17}
D.~Bonn, M.~M. Denn, L.~Berthier, T.~Divoux, S.~Manneville,
  \href{https://link.aps.org/doi/10.1103/RevModPhys.89.035005}{Yield stress
  materials in soft condensed matter}, Rev. Mod. Phys. 89 (2017) 035005.
\newblock \href {https://doi.org/10.1103/RevModPhys.89.035005}
  {\path{doi:10.1103/RevModPhys.89.035005}}.
\newline\urlprefix\url{https://link.aps.org/doi/10.1103/RevModPhys.89.035005}

\bibitem{Kilpatrick12}
P.~K. Kilpatrick, Water-in-crude oil emulsion stabilization: review and
  unanswered questions, Energy \& Fuels 26~(7) (2012) 4017--4026.

\bibitem{Kale17}
S.~N. Kale, S.~L. Deore, Emulsion micro emulsion and nano emulsion: a review,
  Systematic Reviews in Pharmacy 8~(1) (2017) 39.

\bibitem{Mcclements18}
D.~J. McClements, S.~M. Jafari, Improving emulsion formation, stability and
  performance using mixed emulsifiers: A review, Advances in colloid and
  interface science 251 (2018) 55--79.

\bibitem{Rahimian10}
A.~Rahimian, S.~K. Veerapaneni, G.~Biros, Dynamic simulation of locally
  inextensible vesicles suspended in an arbitrary two-dimensional domain, a
  boundary integral method, Journal of Computational Physics 229~(18) (2010)
  6466--6484.

\bibitem{Yang15}
W.~Yang, Z.~Zhou, A.~Yu, D.~Pinson, Particle scale simulation of
  softening--melting behaviour of multiple layers of particles in a blast
  furnace cohesive zone, Powder Technology 279 (2015) 134--145.

\bibitem{Kroupa16}
M.~Kroupa, M.~Vonka, M.~Soos, J.~Kosek, Utilizing the discrete element method
  for the modeling of viscosity in concentrated suspensions, Langmuir 32~(33)
  (2016) 8451--8460.

\bibitem{Seth11}
J.~R. Seth, L.~Mohan, C.~Locatelli-Champagne, M.~Cloitre, R.~T. Bonnecaze, A
  micromechanical model to predict the flow of soft particle glasses, Nature
  materials 10~(11) (2011) 838--843.

\bibitem{Mansard13}
V.~Mansard, A.~Colin, P.~Chaudhuri, L.~Bocquet,
  \href{http://dx.doi.org/10.1039/C3SM50847A}{A molecular dynamics study of
  non-local effects in the flow of soft jammed particles}, Soft Matter 9 (2013)
  7489--7500.
\newblock \href {https://doi.org/10.1039/C3SM50847A}
  {\path{doi:10.1039/C3SM50847A}}.
\newline\urlprefix\url{http://dx.doi.org/10.1039/C3SM50847A}

\bibitem{Jung21}
G.~Jung, S.~M. Fielding, Wall slip and bulk yielding in soft particle
  suspensions, Journal of Rheology 65~(2) (2021) 199--212.

\bibitem{Dollet15}
B.~Dollet, A.~Scagliarini, M.~Sbragaglia,
  \href{https://www.cambridge.org/core/article/two-dimensional-plastic-flow-of-foams-and-emulsions-in-a-channel-experiments-and-lattice-boltzmann-simulations/A8E06679EA25FD323CEDD1695246EF8B}{Two-dimensional
  plastic flow of foams and emulsions in a channel: experiments and lattice
  boltzmann simulations}, J. Fluid Mech. 766 (2015) 556--589.
\newblock \href {https://doi.org/10.1017/jfm.2015.28}
  {\path{doi:10.1017/jfm.2015.28}}.
\newline\urlprefix\url{https://www.cambridge.org/core/article/two-dimensional-plastic-flow-of-foams-and-emulsions-in-a-channel-experiments-and-lattice-boltzmann-simulations/A8E06679EA25FD323CEDD1695246EF8B}

\bibitem{Derzsi17}
L.~Derzsi, D.~Filippi, G.~Mistura, M.~Pierno, M.~Lulli, M.~Sbragaglia,
  M.~Bernaschi, P.~Garstecki, Fluidization and wall slip of soft glassy
  materials by controlled surface roughness, Phys. Rev. E 95~(5) (2017) 052602.

\bibitem{Derzsi18}
L.~Derzsi, D.~Filippi, M.~Lulli, G.~Mistura, M.~Bernaschi, P.~Garstecki,
  M.~Sbragaglia, M.~Pierno, Wall fluidization in two acts: from stiff to soft
  roughness, Soft matter 14~(7) (2018) 1088--1093.

\bibitem{Kruger17}
T.~Kr{\"u}ger, H.~Kusumaatmaja, A.~Kuzmin, O.~Shardt, G.~Silva, E.~M. Viggen,
  The lattice boltzmann method, Springer International Publishing 10~(978-3)
  (2017) 4--15.

\bibitem{Succi18}
S.~Succi, The lattice Boltzmann Equation, Oxford University Press, 2018.

\bibitem{Ludwig01}
J.-C. Desplat, I.~Pagonabarraga, P.~Bladon, Ludwig: A parallel
  lattice-boltzmann code for complex fluids, Computer Physics Communications
  134~(3) (2001) 273--290.

\bibitem{LB3D_2017}
S.~Schmieschek, L.~Shamardin, S.~Frijters, T.~Kr{\"u}ger, U.~D. Schiller,
  J.~Harting, P.~V. Coveney, Lb3d: A parallel implementation of the
  lattice-boltzmann method for simulation of interacting amphiphilic fluids,
  Computer Physics Communications 217 (2017) 149--161.

\bibitem{LBsoft20}
F.~Bonaccorso, A.~Montessori, A.~Tiribocchi, G.~Amati, M.~Bernaschi,
  M.~Lauricella, S.~Succi, Lbsoft: A parallel open-source software for
  simulation of colloidal systems, Computer Physics Communications 256 (2020)
  107455.

\bibitem{Palabos21}
J.~Latt, O.~Malaspinas, D.~Kontaxakis, A.~Parmigiani, D.~Lagrava, F.~Brogi,
  M.~B. Belgacem, Y.~Thorimbert, S.~Leclaire, S.~Li, et~al., Palabos: parallel
  lattice boltzmann solver, Computers \& Mathematics with Applications 81
  (2021) 334--350.

\bibitem{LBfoam21}
M.~Ataei, V.~Shaayegan, F.~Costa, S.~Han, C.~B. Park, M.~Bussmann, Lbfoam: An
  open-source software package for the simulation of foaming using the lattice
  boltzmann method, Computer Physics Communications 259 (2021) 107698.

\bibitem{Moore73}
D.~Moore, N.~Weiss, Two-dimensional rayleigh-b{\'e}nard convection, Journal of
  Fluid Mechanics 58~(2) (1973) 289--312.

\bibitem{Benard1900}
H.~B{\'e}nard, Les tourbillons cellulaires dans une nappe liquide, Rev. Gen.
  Sci. Pures Appl. 11 (1900) 1261--1271.

\bibitem{Rayleigh1916}
L.~Rayleigh, Lix. on convection currents in a horizontal layer of fluid, when
  the higher temperature is on the under side, The London, Edinburgh, and
  Dublin Philosophical Magazine and Journal of Science 32~(192) (1916)
  529--546.

\bibitem{Lohse10}
D.~Lohse, K.-Q. Xia, Small-scale properties of turbulent rayleigh-b{\'e}nard
  convection, Annual Review of Fluid Mechanics 42 (2010) 335--364.

\bibitem{PelusiSM21}
F.~Pelusi, M.~Sbragaglia, R.~Benzi, A.~Scagliarini, M.~Bernaschi, S.~Succi,
  \href{http://dx.doi.org/10.1039/D0SM01777A}{Rayleigh–bénard convection of
  a model emulsion: anomalous heat-flux fluctuations and finite-size droplet
  effects}, Soft Matter 17 (2021) 3709--3721.
\newblock \href {https://doi.org/10.1039/D0SM01777A}
  {\path{doi:10.1039/D0SM01777A}}.
\newline\urlprefix\url{http://dx.doi.org/10.1039/D0SM01777A}

\bibitem{Benzietal09}
R.~Benzi, M.~Sbragaglia, S.~Succi, M.~Bernaschi, S.~Chibbaro,
  \href{http://scitation.aip.org/content/aip/journal/jcp/131/10/10.1063/1.3216105}{Mesoscopic
  lattice boltzmann modeling of soft-glassy systems: Theory and simulations},
  J. Chem. Phys. 131~(10) (2009).
\newblock \href {https://doi.org/http://dx.doi.org/10.1063/1.3216105}
  {\path{doi:http://dx.doi.org/10.1063/1.3216105}}.
\newline\urlprefix\url{http://scitation.aip.org/content/aip/journal/jcp/131/10/10.1063/1.3216105}

\bibitem{Sbragagliaetal12}
M.~Sbragaglia, R.~Benzi, M.~Bernaschi, S.~Succi,
  \href{http://dx.doi.org/10.1039/C2SM26167G}{The emergence of supramolecular
  forces from lattice kinetic models of non-ideal fluids: applications to the
  rheology of soft glassy materials}, Soft Matter 8 (2012) 10773--10782.
\newblock \href {https://doi.org/10.1039/C2SM26167G}
  {\path{doi:10.1039/C2SM26167G}}.
\newline\urlprefix\url{http://dx.doi.org/10.1039/C2SM26167G}

\bibitem{Benzietal14}
R.~Benzi, M.~Sbragaglia, P.~Perlekar, M.~Bernaschi, S.~Succi, F.~Toschi,
  \href{http://dx.doi.org/10.1039/C4SM00348A}{Direct evidence of plastic events
  and dynamic heterogeneities in soft-glasses}, Soft Matter 10 (2014)
  4615--4624.
\newblock \href {https://doi.org/10.1039/C4SM00348A}
  {\path{doi:10.1039/C4SM00348A}}.
\newline\urlprefix\url{http://dx.doi.org/10.1039/C4SM00348A}

\bibitem{LinLin18}
L.~Fei, A.~Scagliarini, A.~Montessori, M.~Lauricella, S.~Succi, K.~H. Luo,
  Mesoscopic model for soft flowing systems with tunable viscosity ratio,
  Physical Review Fluids 3~(10) (2018) 104304.

\bibitem{PelusiSM19}
F.~Pelusi, M.~Sbragaglia, R.~Benzi, Avalanche statistics during coarsening
  dynamics, Soft Matter 15~(22) (2019) 4518--4524.

\bibitem{PelusiEPL19}
F.~Pelusi, M.~Sbragaglia, A.~Scagliarini, M.~Lulli, M.~Bernaschi, S.~Succi, On
  the impact of controlled wall roughness shape on the flow of a soft material,
  {EPL} (Europhysics Letters) 127~(3) (2019) 34005.
\newblock \href {https://doi.org/10.1209/0295-5075/127/34005}
  {\path{doi:10.1209/0295-5075/127/34005}}.

\bibitem{BGK54}
P.~L. Bhatnagar, E.~P. Gross, M.~Krook, A model for collision processes in
  gases. i. small amplitude processes in charged and neutral one-component
  systems, Physical review 94~(3) (1954) 511.

\bibitem{ShanChen93}
X.~Shan, H.~Chen, Lattice boltzmann model for simulating flows with multiple
  phases and components, Physical review E 47~(3) (1993) 1815.

\bibitem{Shan94}
X.~Shan, H.~Chen, Simulation of nonideal gases and liquid-gas phase transitions
  by the lattice boltzmann equation, Physical Review E 49~(4) (1994) 2941.

\bibitem{Falcucci07}
G.~Falcucci, G.~Bella, G.~Chiatti, S.~Chibbaro, M.~Sbragaglia, S.~Succi,
  et~al., Lattice boltzmann models with mid-range interactions, Communications
  in computational physics 2~(6) (2007) 1071--1084.

\bibitem{Falcucci10}
G.~Falcucci, S.~Ubertini, S.~Succi, Lattice boltzmann simulations of
  phase-separating flows at large density ratios: the case of doubly-attractive
  pseudo-potentials, Soft Matter 6~(18) (2010) 4357--4365.

\bibitem{Sbragaglia07}
M.~Sbragaglia, R.~Benzi, L.~Biferale, S.~Succi, K.~Sugiyama, F.~Toschi,
  Generalized lattice boltzmann method with multirange pseudopotential,
  Physical Review E 75~(2) (2007) 026702.

\bibitem{Benzietal15}
R.~Benzi, M.~Sbragaglia, A.~Scagliarini, P.~Perlekar, M.~Bernaschi, S.~Succi,
  F.~Toschi, \href{http://dx.doi.org/10.1039/C4SM02341B}{Internal dynamics and
  activated processes in soft-glassy materials}, Soft Matter 11 (2015)
  1271--1280.
\newblock \href {https://doi.org/10.1039/C4SM02341B}
  {\path{doi:10.1039/C4SM02341B}}.
\newline\urlprefix\url{http://dx.doi.org/10.1039/C4SM02341B}

\bibitem{Ripesi14}
P.~Ripesi, L.~Biferale, M.~Sbragaglia, A.~Wirth, Natural convection with mixed
  insulating and conducting boundary conditions: low-and high-rayleigh-number
  regimes, Journal of fluid mechanics 742 (2014) 636--663.

\bibitem{Delaunay_1934}
B.~Delaunay, \href{http://mi.mathnet.ru/eng/izv4937}{Sur la sph\`ere vide},
  Bulletin de l'Acad\'emie des Sciences de l'URSS, Classe des sciences
  math\'ematiques et naturelles 6 (1934).
\newline\urlprefix\url{http://mi.mathnet.ru/eng/izv4937}

\bibitem{Bernaschi16}
M.~Bernaschi, M.~Lulli, M.~Sbragaglia,
  \href{http://www.sciencedirect.com/science/article/pii/S0010465516303599}{{GPU}
  based detection of topological changes in voronoi diagrams}, Comput. Phys.
  Commun. 213 (2017) 19 -- 28.
\newblock \href {https://doi.org/http://dx.doi.org/10.1016/j.cpc.2016.11.005}
  {\path{doi:http://dx.doi.org/10.1016/j.cpc.2016.11.005}}.
\newline\urlprefix\url{http://www.sciencedirect.com/science/article/pii/S0010465516303599}

\bibitem{Swendsen_1987}
R.~H. Swendsen, J.-S. Wang,
  \href{http://link.aps.org/doi/10.1103/PhysRevLett.58.86%5Cnhttp://kobus.ca/teaching/cs645/spring09/ua_only/swendsen-wang-87.pdf
  https://link.aps.org/doi/10.1103/PhysRevLett.58.86}{{Nonuniversal critical
  dynamics in Monte Carlo simulations}}, Physical Review Letters 58~(2) (1987)
  86--88.
\newblock \href {https://doi.org/10.1103/PhysRevLett.58.86}
  {\path{doi:10.1103/PhysRevLett.58.86}}.
\newline\urlprefix\url{http://link.aps.org/doi/10.1103/PhysRevLett.58.86%5Cnhttp://kobus.ca/teaching/cs645/spring09/ua_only/swendsen-wang-87.pdf
  https://link.aps.org/doi/10.1103/PhysRevLett.58.86}

\bibitem{Komura_2012}
Y.~Komura, Y.~Okabe, \href{http://dx.doi.org/10.1016/j.cpc.2012.01.017
  https://linkinghub.elsevier.com/retrieve/pii/S001046551200032X}{{GPU-based
  Swendsen–Wang multi-cluster algorithm for the simulation of two-dimensional
  classical spin systems}}, Computer Physics Communications 183~(6) (2012)
  1155--1161.
\newblock \href {http://arxiv.org/abs/1202.0635} {\path{arXiv:1202.0635}},
  \href {https://doi.org/10.1016/j.cpc.2012.01.017}
  {\path{doi:10.1016/j.cpc.2012.01.017}}.
\newline\urlprefix\url{http://dx.doi.org/10.1016/j.cpc.2012.01.017
  https://linkinghub.elsevier.com/retrieve/pii/S001046551200032X}

\bibitem{Voronoi_1908}
G.~Voronoi,
  \href{https://www.degruyter.com/document/doi/10.1515/crll.1908.134.198/html}{{Nouvelles
  applications des param{\`{e}}tres continus {\`{a}} la th{\'{e}}orie des
  formes quadratiques. Deuxi{\`{e}}me m{\'{e}}moire. Recherches sur les
  parall{\'{e}}llo{\`{e}}dres primitifs.}}, Journal f{\"{u}}r die reine und
  angewandte Mathematik (Crelles Journal) 1908~(134) (1908) 198--287.
\newblock \href {https://doi.org/10.1515/crll.1908.134.198}
  {\path{doi:10.1515/crll.1908.134.198}}.
\newline\urlprefix\url{https://www.degruyter.com/document/doi/10.1515/crll.1908.134.198/html}

\bibitem{Januszewski_2014}
M.~Januszewski, M.~Kostur,
  \href{https://doi.org/10.1016%2Fj.cpc.2014.04.018}{Sailfish: A flexible
  multi-{GPU} implementation of the lattice boltzmann method}, Computer Physics
  Communications 185~(9) (2014) 2350--2368.
\newblock \href {https://doi.org/10.1016/j.cpc.2014.04.018}
  {\path{doi:10.1016/j.cpc.2014.04.018}}.
\newline\urlprefix\url{https://doi.org/10.1016%2Fj.cpc.2014.04.018}

\bibitem{BernaschiGPU09}
M.~Bernaschi, L.~Rossi, R.~Benzi, M.~Sbragaglia, S.~Succi,
  \href{http://link.aps.org/doi/10.1103/PhysRevE.80.066707}{Graphics processing
  unit implementation of lattice boltzmann models for flowing soft systems},
  Phys. Rev. E 80 (2009) 066707.
\newblock \href {https://doi.org/10.1103/PhysRevE.80.066707}
  {\path{doi:10.1103/PhysRevE.80.066707}}.
\newline\urlprefix\url{http://link.aps.org/doi/10.1103/PhysRevE.80.066707}

\bibitem{Ahrens_2014}
J.~Ahrens, S.~Jourdain, P.~O{\textquotesingle}Leary, J.~Patchett, D.~H. Rogers,
  M.~Petersen, \href{https://doi.org/10.1109%2Fsc.2014.40}{An image-based
  approach to extreme scale in situ visualization and analysis}, in: {SC}14:
  International Conference for High Performance Computing, Networking, Storage
  and Analysis, {IEEE}, 2014.
\newblock \href {https://doi.org/10.1109/sc.2014.40}
  {\path{doi:10.1109/sc.2014.40}}.
\newline\urlprefix\url{https://doi.org/10.1109%2Fsc.2014.40}

\bibitem{Pal96}
R.~Pal, Effect of droplet size on the rheology of emulsions, AIChE Journal
  42~(11) (1996) 3181--3190.

\bibitem{Zhang06}
J.~Zhang, D.~Vola, I.~A. Frigaard, Yield stress effects on rayleigh–benard
  convection, J. Fluid Mech. 566 (2006) 389--419.
\newblock \href {https://doi.org/10.1017/S002211200600200X}
  {\path{doi:10.1017/S002211200600200X}}.

\bibitem{Shraiman90}
B.~I. Shraiman, E.~D. Siggia, Heat transport in high-rayleigh-number
  convection, Physical Review A 42~(6) (1990) 3650.

\bibitem{Ahlers09}
G.~Ahlers, S.~Grossmann, D.~Lohse, Heat transfer and large scale dynamics in
  turbulent rayleigh-b{\'e}nard convection, Reviews of modern physics 81~(2)
  (2009) 503.

\bibitem{Verzicco10}
R.~J. Stevens, R.~Verzicco, D.~Lohse, Radial boundary layer structure and
  nusselt number in rayleigh-b{\'e}nard convection, Journal of Fluid Mechanics
  643 (2010) 495--507.

\bibitem{Chilla12}
F.~Chill{\`a}, J.~Schumacher, New perspectives in turbulent rayleigh-b{\'e}nard
  convection, The European Physical Journal E 35~(7) (2012) 1--25.

\end{thebibliography}







\end{document}